\documentclass[%
 reprint,
 amsmath,amssymb,
 aps,
]{revtex4-2}

\usepackage{graphicx}
\usepackage{dcolumn}
\usepackage{bm}

\usepackage{float}

\begin{document}

\preprint{APS/123-QED}

\title{Chaotic Dynamics due to Prolate and Oblate Sources in Kerr-like and Hartle-Thorne Spacetimes with and without Magnetic Field}

\author{Adri\'an Eduarte-Rojas}
\email{adrian.eduarte@ucr.ac.cr}
\affiliation{Space Research Center (CINESPA) \\ School of Physics \\ 
University of Costa Rica}

\author{Francisco Frutos-Alfaro}%
\email{francisco.frutos@ucr.ac.cr}
\homepage{http://www.cinespa.ucr.ac.cr/}
\affiliation{Space Research Center (CINESPA) \\ School of Physics \\ 
University of Costa Rica}

\author{Rodrigo Carboni}%
\email{rodrigo.carboni@ucr.ac.cr}
\affiliation{Space Research Center (CINESPA) \\ School of Physics \\ 
University of Costa Rica}




\date{\today}

\begin{abstract}

As demonstrated by observations, every stellar-mass object rotates around some axis; some objects spin faster than others due to different mechanisms. Furthermore, these spinning objects are slightly deformed and are no longer perfect spheres because of hydrostatic equilibrium. The well-known Kerr solution of the Einstein Field Equations (EFE) represents the spacetime surrounding a rotating spherical gravitational source. However, real objects deviate from a perfect sphere and may be prolate or oblate. There are several solutions of the EFE that represent the spacetime of deformed objects. The Kerr--like (KL) metric represents the spacetime surrounding this kind of object, where the deformation is characterized by the mass quadrupole moment parameter $q_{\mathrm{KL}}$. When $q_{\mathrm{KL}} \neq 0$, the Carter constant no longer exists and the equations of motion (EOM) are no longer integrable; therefore, the system exhibits chaotic orbits. Another widely used solution is the Hartle--Thorne (HT) metric, which has similar characteristics and represents a slightly deformed, slowly rotating star. The HT metric has several versions, and two of them were selected to test their validity. The traditional HT version, which contains logarithmic terms, is less accurate than the version with exponential terms. Moreover, both the KL and HT metrics may be extended to include contributions due to the magnetic dipole moment of the source. The equations of motion (EOM) were computed, and these new dynamical systems display several interesting features, which are shown in their Poincar'e sections.

In this contribution, we compare the dynamics of a unit-mass test particle under the influence of the Kerr-like and Hartle--Thorne spacetimes for prolate gravitational sources ($q_{\mathrm{KL}} > 0$) and oblate ones ($q_{\mathrm{KL}} < 0$), as well as their extensions including a magnetic dipole moment $\mu_d$.

\end{abstract}

\maketitle


\section{\label{sec:Intro}Introduction}

Astrophysical compact objects such as neutron stars, white dwarfs, black holes, or even planets exhibit a certain degree of rotation. As demonstrated by Kerr, the spacetime surrounding a rotating gravitational source presents phenomena that are not present in non-spinning objects, such as frame dragging. For very slow rotation, this effect may be neglected; however, as observations of these stars improve, the number of highly spinning objects increases. For example, millisecond pulsars are among the fastest rotating objects known. New pulsars are discovered every year, and astronomers often report findings that push the limits of our current understanding of the mechanics and behavior of neutron stars \cite{MSPgamma}.

The Kerr metric is an idealization of the surrounding spacetime of stars because rotation can deform the object; similarly, interactions with other massive objects may produce the same effect. There is no perfectly rigid body in the Universe. The Kerr metric represents only a perfectly massive sphere and has another limitation: it cannot be extended to an interior solution \cite{QuevedoCompactObj}. Apart from rotating and uncharged black holes, the Kerr metric is unsuitable for nearly every other type of object, from planets to stars and their remnants, such as white dwarfs and neutron stars.

On the other hand, a widely used approximate solution of the EFE, known as the Hartle--Thorne metric (HT), may be used to describe the exterior of the gravitational source, and it could be used to construct interior solutions with physical relevance \cite{QuevedoHTinterior}. The advantage of the HT metric is that it includes the mass quadrupole moment $q_{HT}$ and the rotation parameter $a$. Compact objects such as neutron stars are studied using this metric in the regime of slow rotation. There are many versions of the HT metric; some of them correspond to linear order in rotation and mass quadrupole moment, while others include terms up to second order in these parameters. There is extensive work on the HT metric showing chaotic motion \cite{chaosHTKyriakos, misbah}, which may be measured through observations of gravitational waves. The HT metric is one of the favored spacetime models due to its simplicity for numerical work. The main issue with the HT metric is the low order of its multipole moments; moreover, the presence of complicated functions, such as associated Legendre functions of the second kind, reduces this advantage of simplicity.

In a previous contribution \cite{adrian}, the properties of the dynamics of a test particle under the Kerr-like metric with Mass Quadrupole Moment (KL) were studied. This metric is an approximate solution of the EFE, found using the Ernst formalism \cite{Carmeli} and the Hoenselaers--Kinnersley--Xanthopoulos (HKX) transformations \cite{frutosmetric}. The KL metric may be considered an improvement over the HT spacetime, since it includes the spin octupole moment, which is absent in the HT metric \cite{frutosmetric}. Numerical computations are easy to perform using this spacetime model. Although the KL metric has a more complex structure in the expressions for its metric components than the HT metric, it does not involve associated Legendre functions of the second kind, but only Legendre polynomials of the first kind for the angular coordinate. Therefore, the KL metric is expected to be more accurate and less time-consuming than the HT metric.

For a general metric representing a stationary spacetime surrounding a rotating body

\begin{equation}
\label{eq:mgen}
ds^2 = - Vdt^2 + 2Wdtd\phi +Xdr^2 +Yd\theta^2 +Zd\phi^2 , 
\end{equation}

\noindent
where the potentials $V$, $W$, $X$, $Y$, and $Z$ are functions of $r$ and $\theta$ only. This general metric covers a family of stationary solutions of the EFE for the exterior of the gravitational source. For example, the Kerr metric \cite{gravitationmisnerthorne}, the Manko--Novikov metric \cite{contopoulosgerakopoulos}, and the Quevedo--Mashoon metric \cite{QuevedoMashoon}, which are exact solutions. Moreover, the HT and KL metrics are approximate solutions of the EFE. Although exact solutions are clearly an improvement over the Kerr metric for different types of gravitational sources due to their multipole parameters, they are inconvenient for numerical work. The approximate solutions present high accuracy, fewer free parameters, and lower computational cost.

The equations of motion for a test particle with mass $\mu$ are obtained using the Hamiltonian formalism \cite{gravitationmisnerthorne, adrian}

\begin{equation}
\begin{split}
p_t &= \mu ( -V \dot{t} +W\dot{\phi})= -E , \\
p_r &= \mu X \dot{r} ,  \\
p_\theta &= \mu Y\dot{\theta} ,   \\
p_\phi &= \mu (W\dot{t} + Z \dot{\phi}) = L_z .
\end{split}
\end{equation}

The Lagrangian does not depend directly on $t$ and $\phi$ because neither does the metric potential; therefore, their respective conjugate momenta are constants of motion corresponding to the orbital energy $p_t = -E$ and the orbital angular momentum $p_\phi = L_z$. The Hamiltonian is also a constant of motion, and its value may be obtained using the energy-momentum relation $p_\mu p^\mu = -\mu^2$. Then, $\mathcal{E} = -\mu$ (in units where $c = 1$) \cite{adrian}.

\begin{equation} \label{E:condini}
\mathcal{H} = \frac{\mu}{2} \left[\frac{p_r ^2}{\mu^2 X} + \frac{p_\theta^2}{\mu^2 Y} + V_{\rm eff}\right] = \mathcal{E} ,
\end{equation}

\noindent
where $ \tilde{\rho}^2 = VZ+W^2$, and the effective potential is given by 

\begin{equation}\label{E:Veff}
    V_{\rm eff} = - \frac{1}{\mu^2 \tilde{\rho}^2}{(E^2 Z + 2EL_z W -L_z ^2 V)} .
\end{equation}

Equation (\ref{E:condini}) constrains the momenta $p_r$ and $p_{\theta}$. The canonical equations of motion correspond to the dynamics of the phase space of the system and are given by \cite{adrian, gravitationmisnerthorne}

\begin{equation} 
\label{E:ecmov}
    \begin{split}
        \dot{r} &= \frac{p_r}{\mu X} , \\
        \dot{\theta} &= \frac{p_\theta}{\mu Y} , \\
        \dot{t} &= \frac{1}{ \mu \tilde{\rho}^2}( EZ + L_z W ) ,  \\
        \dot{\phi} &= \frac{1}{\mu  \tilde{\rho}^2}( L_z V - EW ) , \\
        \dot{p_r} &= \frac{1}{2} \left(  \frac{p_r^2}{X^2} \partial_r X + \frac{p_\theta^2}{Y^2} \partial_r Y  - \partial_r V_{eff} \right) , \\
        \dot{p_\theta} &= \frac{1}{2} \left(  \frac{p_r ^2}{X^2} \partial_\theta X  + \frac{p_\theta ^2}{Y^2} \partial_\theta Y - \partial_\theta V_{eff} \right) , \\
        \dot{p_\phi} &= 0  , \\
        \dot{p_t} &= 0 .
    \end{split}
\end{equation}

To find the solution, it is important to take into account that only one initial condition for the momenta is needed, because Eq.~(\ref{E:condini}) constrains the radial momentum $p_r$ with the angular momentum $p_\theta$. To simplify the numerical solutions, we take the test particle mass to be $\mu = 1$ and later select a collection of parameters and initial conditions \cite{adrian}.

Finally, in the geodesic simulations, the Boyer--Lyndquist coordinates are used \cite{adrian}

\begin{eqnarray}
        x &=& \sqrt{r^2 + a^2} \sin{\theta} \cos{\phi}, \nonumber\\
        y &=& \sqrt{r^2 + a^2} \sin{\theta} \sin{\phi}, \\
        z &=& r \cos{\theta}.   \nonumber
\end{eqnarray}

\section{Test metrics}

As stated in the previous section, there are several solutions of the EFE that are of the form (\ref{eq:mgen}). Some of these are exact solutions, such as Kerr, Quevedo--Mashoon, and Manko--Novikov, while others are approximate solutions, such as KL and HT. The advantage of approximate solutions is that their convergence rate is high and they require less computational time than exact solutions due to their simplified forms.

The KL metric takes the following form \cite{frutosmetric, frutos2}

\begin{eqnarray}\label{E:KLmetric}
V &=& \frac{{\rm e}^{-2\psi}}{\rho^2} (\Delta -a^2 \sin ^2 \theta ) ,
\nonumber \\
W &=& -\frac{2Mar}{\rho^2}\sin^2 \theta , \nonumber \\
X &=& \rho^2 \frac{{\rm e}^{2\chi}}{\Delta}, \\
Y &=& \rho^2 {\rm e}^{2\chi} , \nonumber \\
Z &=&  \frac{{\rm e}^{2\psi}}{\rho^2}( (r^2 +a^2)^2 - a^2\Delta \sin ^2\theta)
\sin^2\theta , \nonumber 
\end{eqnarray}

\noindent
where
 
\begin{equation}
\psi = \frac{q_{KL}}{r^3}P_2 + 3 \frac{Mq_{KL}}{r^4}P_2 ,
\end{equation}

\begin{equation}
\begin{split}
\chi = \frac{q_{KL} P_2}{r^3} + \frac{1}{3}\frac{Mq}{r^4} (-1 + 5 P_2 + 5 P_2 ^2) \\ + \frac{1}{9} \frac{q_{KL}^2 }{r^6}(2 + 6 P_2 + 21 P_2 ^2 + 25 P_2 ^3) ,    
\end{split}
\end{equation}

\begin{equation}
    \rho^2 = r^2 + a^2 \cos ^2 \theta,
\end{equation}

\begin{equation} \label{E:Delta}
    \Delta = r^2 -2 M r + a^2,
\end{equation}

\noindent
and $ P_2 $ is a Legendre polynomial ($P_2 = {(3\cos^2{\theta}  -1)/2}$).

The KL is an approximate solution of the EFE for a rotating compact object. Given a seed metric, in this case the original Kerr metric, perturbation terms are later added \cite{frutosmetric, frutos2}. A test particle in the Kerr system presents integrable equations of motion due to spherical symmetry; it has four constants of motion:  the orbital energy $E$, the $z$-component of the angular momentum $L_z$, the rest mass of the test particle $\mu$, which is related to the rest energy $\varepsilon = -\mu$, and the Carter constant $\mathfrak{C}$ \cite{gravitationmisnerthorne}. However, when it is perturbed by a mass quadrupole moment $q_{KL}$, the Carter constant is lost and the equations of motion are no longer integrable. The phase space of integrable systems shows only the main island of stability, whereas non-integrable systems exhibit several structures in the regions where the gravitational field is stronger. The Kerr-like metric with $q_{KL} \ne 0$ presents chaotic regions; hence, it is a non-integrable system \cite{adrian}. The lack of the Carter constant is also present in other axisymmetric, stationary metrics, such as the Manko--Novikov metric or the Zipoy--Vorhees metric, giving results similar to those presented in this work~\cite{countopouloslukes2014non, HowtoObserveNonKerr, LukesZpoyVoorhees}. However, these metrics contain several complicated factors that make them unsuitable for numerical computation.

On the other hand, the Hartle--Thorne metric is a widely used approximate solution of the EFE for the study of spacetime, both interior and exterior, of slowly rotating and slightly deformed compact objects \cite{ERvsHT}. As with the KL metric, the Hartle--Thorne metric is determined by the source mass $M$, the spin parameter $a$, and the quadrupole parameter $q_{HT}$.

There are several versions of the HT metric; the first one considered here is the logarithmic version (HTlog), whose potentials are given by \cite{ERvsHT, frutosmetric}

\begin{eqnarray}\label{E:HT}
V &=& (1 - 2U)[1 + 2K_1 P_2 (\cos \theta)]  + \frac{2J^2}{r^4} ( 2\cos^2 \theta   -1 ) ,     \nonumber \\
W &=& -\frac{2J}{r}\sin^2 \theta , \nonumber \\
X &=&  \frac{1}{1 - 2U} \left[ 1 - 2K_2 P_2 (\cos \theta)  - \frac{2}{1 - 2U} \frac{J^2}{r^4}  \right] ,  \\
Y &=&  r^2 [ 1 - 2K_3 P_2 (\cos \theta)  ]      , \nonumber \\
Z &=&  Y\sin^2\theta , \nonumber 
\end{eqnarray}

\noindent
where

\begin{eqnarray}\label{E:HTKs}
K_1 &=& \frac{J^2}{Mr^3}(1 + U)  + \frac{5}{8}  \left(  \frac{q_{HT}}{M^3}  \right. \nonumber \\
&-&  \left.   \frac{J^2}{M^4}    \right) Q^2 _2  \left(  \frac{r}{M}  - 1  \right)   , \nonumber \\
K_2 &=& K_1 - \frac{6J^2}{r^4}            ,  \\
K_3 &=& \left( K_1  + \frac{J^2}{r^4}  \right) +  \frac{5}{4}  \left(  \frac{q_{HT}}{M^3}  \right. \nonumber \\
&-&  \left. \frac{J^2}{M^4}    \right)  \frac{U}{ \sqrt{1 - 2U}  }   Q^1 _2  \left(  \frac{r}{M}  - 1  \right)    , \nonumber
\end{eqnarray}

\begin{equation}
U = \frac{M}{r} 
\end{equation}

The functions $Q_2  ^1 (x)$ and $Q_2 ^2 (x)$ are the associated Legendre functions of the second kind

\begin{eqnarray}\label{E:legendre2kind}
Q^1 _2 &=&  \sqrt{x^2 -1} \left( \frac{3}{2}x \ln  \left[  \frac{x + 1}{x -1}  \right]   - \frac{3x^2 - 2}{x^2 -1}       \right)     , \nonumber \\
Q^2 _2 &=&   (x^2 -1)  \left( \frac{3}{2} \ln  \left[  \frac{x + 1}{x - 1}  \right]   - \frac{3x^3 - 5x}{(x^2 -1)^2}       \right)     .           
\end{eqnarray}

\noindent
and $x = (r/M) - 1$.

Again, this metric does not depend explicitly on $t$ and $\phi$; therefore, the equations of motion (\ref{E:ecmov}) are the same. This spacetime is also compatible with interior solutions \cite{ERvsHT}. In addition, the HTlog metric, as well as the KL metric, shows structures in phase space due to the non-integrability of the system \cite{misbah}.

The HT and KL metrics are not isometric; hence, they do not represent the same spacetime with their respective parameters of mass, spin, and mass quadrupole moment. In order to represent the same object with these selected spacetime models, a direct relation between their Geroch--Hansen multipole moments is needed. These multipole moments are coordinate independent and carry information about the spacetime outside an arbitrary mass distribution \cite{geroch}. Since the two metrics studied in this paper represent the exterior spacetime of a spheroid with mass quadrupole moment, it is only necessary to consider terms up to second order in both metrics. The Fodor--Hoenselaers--Perj\'es method is an algorithm used to construct the $n$-th multipole moment of an axisymmetric, stationary, rotating spacetime by using the Ernst potential \cite{fodor, frutosmetric} and a Taylor series expansion. The Hartle--Thorne multipole moments are given by \cite{ERvsHT}; however, only the $\mathcal{M}_2$ moment is required, since it contains the information of the mass quadrupole moment:

\begin{equation}
    \mathcal{M} _2 = -q_{HT}
\end{equation}

\noindent
and for the Kerr--like metric \cite{frutosmetric}

\begin{equation}
    \mathcal{M} _2 = q_{KL} - Ma^2
\end{equation}

By comparing the mass moment $\mathcal{M} _2$, there is a correspondence between the mass quadrupole moment of the Kerr--like metric $q_{KL}$ and the equation \cite{frutosmetric}

\begin{equation} \label{E:rel_q}
   q_{HT} = Ma^2 -q_{KL}
\end{equation}

As stated in \cite{frutosmetric}, when both metrics are expanded using a Taylor series and compared using (\ref{E:rel_q}), the Kerr-like metric is found to be an improvement over the Hartle--Thorne metric. Therefore, the Kerr-like metric with mass quadrupole moment up to second order in $a$, $J$, $M$, and $q$ takes the form:

\begin{eqnarray}\label{E:KLmetricTaylor}
    V &=& 1 - 2\frac{M}{r} - 2 \left(  Ma^2 \cos^2 \theta  - q P_2    \right) \frac{1}{r^3}  \nonumber \\ 
    &-& \frac{Mq}{r^4}P_2  + 2 \frac{q^2}{r^6}(P_2 )^2        ,\nonumber \\
    W &=&  -2\frac{Ma}{r}\sin^2 \theta        , \nonumber \\
    X &=&  1 + 2\frac{M}{r} + \left( 4 M^2  -a^2 \sin^2 \theta \right.   \\
    &-&  \left.  2 M a^2 (1 + \sin^2 \theta)    \right)\frac{1}{r^2}  + 2\frac{q}{r^3} P_2 \nonumber \\
    &+&  \bigg( -4M^2 a^2  \left( 2 + \sin^2 \theta   \right)    \nonumber\\
    &+& \frac{2}{3}Mq \left(  5 (P_2 )^2  + 11P_2 -1 \right)  \bigg) \frac{1}{r^4}  \nonumber \\
    &+& \frac{2}{9} \frac{q^2}{r^6} \left(  25 (P_2)^3  - 12(P_2)^2 - 6P_2 + 2 \right),   \nonumber \\
    Y &=&  r^2 \left(  1 + 2\frac{q}{r^3} P_2 \right. \nonumber\\
    &+& \left(a^2 \cos^2 \theta  + \frac{2}{3}Mq \left(5(P_2)^2 + 5 P_2 - 1 \right)   \right) \frac{1}{r^4}      \nonumber \\
    &+& \left. \frac{2}{9}\frac{q^2}{r^6}  \left(25(P_2)^3 -12(P_2)^2 -6 P_2 +2 \right) \right) , \nonumber \\
    Z &=&   r^2  \sin^2 \theta \left(  1 + \frac{a^2}{r^2}  + \left(  2Ma^2 \sin^2 \theta + 2q P_2   \right)\frac{1}{r^3}    \right.    \nonumber \\
    &+& \left.  6\frac{Mq}{r^4}P_2   + 2\frac{q^2}{r^6}(P_2)^2  \right)    . \nonumber 
\end{eqnarray}

\noindent
and the Hartle--Thorne metric takes the form:

\begin{eqnarray}\label{E:HTmetricTaylor}
V &=&   1 - 2\frac{M}{r}  + 2\frac{q}{r^3}P_2  + \Bigl(2MqP_2 \Bigr. \nonumber \\ 
&-& \Bigl. \frac{2}{3} M^2 a^2 (2P_2 + 1) \Bigr)\frac{1}{r^4}  + 2\frac{q^2}{r^6}(P_2)^2  ,\nonumber \\
W &=&  -2\frac{Ma}{r}\sin^2 \theta        , \nonumber \\
X &=& 1 + 2\frac{M}{r} + 2\frac{M^2}{r^2}   -2 \frac{q}{r^3}P_2 + (  2M^2 a^2(8P_2 -1) \nonumber \\
&-& 10 MqP_2   )\frac{1}{r^4}  \nonumber \\
&+&  \frac{1}{12}q^2 \left(  8 (P_2)^2  -16 P_2 +77  \right)\frac{1}{r^6} , \\
Y &=&  r^2 \left(  1 -2\frac{q}{r^3}P_2 + ( M^2 a^2  -5Mq )\frac{P_2}{r^4}     \right.   \nonumber \\
&+&  \left.  \frac{1}{36}\frac{q^2}{r^6}\left(  44(P_2)^2 +8 P_2 -43   \right)  \right)   , \nonumber \\
Z &=&   r^2 \sin^2 \theta \left(  1 -2\frac{q}{r^3}P_2 + ( M^2 a^2  -5Mq )\frac{P_2}{r^4}     \right.   \nonumber \\
&+&  \left.  \frac{1}{36}\frac{q^2}{r^6}\left(  44(P_2)^2 +8 P_2 -43   \right)  \right)       . \nonumber 
\end{eqnarray}

Both metrics coincide when equation (\ref{E:rel_q}) is taken into account \cite{frutosmetric}.

\cite{frutos2} presents another form of the Hartle--Thorne metric, which we call the approximate HT (appHT), in which logarithmic terms are absent, but the metric components contain exponential factors. The metric form is given by:

\begin{eqnarray} \label{E:appHT}
    V_{HT} &\simeq& \left(     1 - 2U - \frac{2}{3}\frac{J^2}{r^4}   \right) e^{2\alpha_1} ,  \nonumber  \\
    W_{HT} &=&      -2 \frac{J}{r} \sin^2 \theta  ,  \nonumber\\
    X_{HT} &\simeq& \left(   1 - 2U + 2 \frac{J^2}{r^4}   \right)^{-1}  e^{-2\alpha_2} ,  \\
    Y_{HT} &\simeq& r^2  e^{-2\alpha_3} ,  \nonumber \\
    Z_{HT} &\simeq& r^2 \sin^2 \theta e^{-2\alpha_3} .  \nonumber 
\end{eqnarray}

where

\begin{eqnarray}
    U = M/r
\end{eqnarray}

\begin{eqnarray}
    \alpha_1 &=&  \left(    K_1 + \frac{4}{3}\frac{J^2}{r^4}  \right)P_2 ,  \nonumber \\
    \alpha_2 &=&  K_2 P_2 , \\
    \alpha_3 &=&  K_3 P_2. \nonumber
\end{eqnarray}

and the $K_i$ are given by Taylor series expansion

\begin{eqnarray}
    K_1 &=&  \frac{q}{r^3} + 3 \frac{mq}{r^4} - 2 \frac{J^2}{r^4} , \nonumber  \\
    K_2 &=&  \frac{q}{r^3} + 3 \frac{mq}{r^4} - 8 \frac{J^2}{r^4}  ,  \\
    K_3 &=&  \frac{q}{r^3} + \frac{5}{2} \frac{mq}{r^4} - \frac{1}{2} \frac{J^2}{r^4}.  \nonumber  \\
\end{eqnarray}

This version of the Hartle--Thorne metric also preserves the relation (\ref{E:rel_q}).

Another related metric is the Quevedo–Mashhoon metric (QM) \cite{QuevedoMashoon}, which represents a deformed compact object through its own mass quadrupole moment, $q_{\mathrm{QM}}$. The QM metric is an exact solution of the EFE with an infinite number of free parameters representing the multipole moments. It can be reduced by retaining only terms up to the quadrupole moment; however, this leads to a complicated set of equations. The numerical integration time for the QM metric is longer than that for the HT or KL metrics. Moreover, the QM metric is not isometric with either KL or HT. Therefore, we do not consider the QM metric in this work.

\begin{figure}[h]
    \centering
    \includegraphics[width=0.4\textwidth]{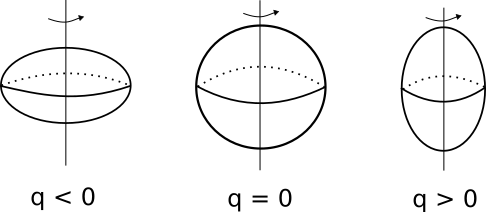}
    \caption{Representation of mass distribution according its mass quadrupole moment $q_{KL}$, for Erez--Rozen metric and Kerr-like metric. The sign is reversed for Hartle--Thorne metric.}    
    \label{F:oblateprolate}
\end{figure}

Figure~\ref{F:oblateprolate} shows how compact objects (sources of the gravitational field) are deformed by the mass quadrupole moment. A positive mass quadrupole moment represents prolate objects, while a negative one represents oblate objects in the Erez--Rosen metric \cite{ERvsHT}. The transformation between the Kerr-like metric and the Erez--Rosen metric preserves the sign of $q$.

An extension of the Kerr-like metric and the Hartle--Thorne metric was developed in \cite{frutosdipolar}, which includes contributions from the magnetic dipole moment $\mu_d$ and the electric charge $q_e$ of the gravitational source. These metrics were obtained using the Ernst formalism and reduce to the Kerr-like (KL) metric and the approximate Hartle--Thorne (appHT) metric, respectively, when the magnetic dipole moment $\mu_d \rightarrow 0$ and $q_e \rightarrow 0$.

There are two versions of the dipolar Kerr-like (dipKL) metric \cite{frutosdipolar}, and the exponential form was selected because it has numerical advantages.

\begin{eqnarray}\label{E:KLdip}
V &=& \frac{{\rm e}^{2(\psi_{1\mu} - \psi_q)}}{\rho^2} (\Delta -a^2 \sin ^2 \theta ) ,
\nonumber \\
W &=& \frac{a}{\rho^2}\left[ \Delta - (r^2 + a^2) \right] \sin^2 (\theta) +  W_q + W_\mu   , \nonumber \\
X &=& \rho^2 \frac{{\rm e}^{2(\chi_q + \chi_{1\mu} ) }}{\Delta}, \\
Y &=& \rho^2 {\rm e}^{2(\chi_q + \chi_{2\mu} )} , \nonumber \\
Z &=&  \frac{{\rm e}^{2(\psi_q + \psi_{2\mu})}}{\rho^2}( (r^2 +a^2)^2 - a^2\Delta \sin ^2\theta)
\sin^2\theta . \nonumber 
\end{eqnarray}

where $\rho^2 = r^2 +a^2 \cos^2 \theta$. The auxiliary functions are

\begin{eqnarray}\label{E:KLdip_aux}
    \psi_q &=& qP_2u^3 + 3MqP_2u^4 , \nonumber   \\
    \psi_{1\mu} &=& \frac{1}{6}\mu^2 _d (2P2 +1)u^4 = \frac{1}{2}\mu^2 _d u^4 \cos(\theta), \nonumber \\
    \psi_{2\mu} &=& \frac{1}{36}\mu^2 _d (19 - 7P_2)u^4, \nonumber  \\
    \chi_q &=&  qP_2 u^3 + \frac{1}{9}Mq(15(P_2) ^2 + 15P_2 -3)u^4  \\
           &+& \frac{1}{9}q^2 (25(P_2) ^3 -21 P_2 ^2 -6 P_2 + 2)u^6 , \nonumber \\
    \chi_{1\mu} &=& \frac{1}{8}\mu^2 _d (4P_2 -15)u^4, \nonumber \\
    \chi_{2\mu} &=& \frac{1}{6}\mu^2 _d (3 - P_2)u^4. \nonumber
\end{eqnarray}

and the contributions to the $W$ metric potential are

\begin{eqnarray}
    W_q &=& -Maq P_3 ^1 u^4 \sin \theta , \nonumber \\
    W_\mu &=& \mu q_e (\beta_3 P_2 + \beta_4)u² \sin^2 \theta. 
\end{eqnarray}

The electromagnetic four-potential components are perturbed as well, and are given by

\begin{eqnarray}\label{fourpotential}
    A_t &=& K_t - \frac{1}{5}qq_e(-5P_2 + 2P_3)u^4 - Ma\mu_d u^4 \cos^2(\theta) ,\nonumber \\
    A_r &=& 0,  \\
    A_\theta &=& 0, \nonumber \\
    A_{\phi} &=& K_{\phi} + \left(  \mu_d u + \frac{3}{2}M\mu_d u^2 - \frac{1}{4} \mu_d q (P_2 +1)u^4 \right)\sin^2(\theta). \nonumber
\end{eqnarray}

where $K_\mu = (K_t , 0, 0, K_\phi)$ is the unperturbed part of the electromagnetic four-potential for the Kerr--Newman metric \cite{gravitationmisnerthorne, frutosdipolar}  

\begin{eqnarray}
    K_t &=& -\frac{q_e r}{\rho^2}, \\
    K_\phi &=&  -a K_t \sin^2 (\theta) .  \nonumber 
\end{eqnarray}

The potentials for the Hartle--Thorne metric with dipolar magnetic contributions (dipHT) are given by

\begin{eqnarray}\label{E:HTdip}
    V &=& \left(    1 - 2\frac{M}{r} + \frac{q_e ^2}{r^2}  - \frac{2}{3}\frac{J^2}{r^4} \right) e^{2\psi_1},  \nonumber \\
    W &=& -2 \frac{J\sin^2 \theta}{r} - \frac{JqP^1 _3 \sin \theta}{r^4} + W_\mu , \nonumber \\
    X &=& \left(  1 - 2 \frac{M}{r}  + \frac{q_e ^2}{r^2}  + 2\frac{J^2}{r^4} \right)^{-1} e^{-2\psi_2}, \\
    Y &=& r^2 e^{-2\psi_3} ,\nonumber  \\
    Z &=& Y \sin^2 \theta. \nonumber
\end{eqnarray}

where $P_3 ^1 = (5P_2 + 1)|\sin \theta|$, although the absolute value is not needed because of the domain $\theta \in [0, \pi]$. The auxiliary functions are:

\begin{eqnarray}
    \psi_1 &=& \frac{qP_2}{r^3} + 3 \frac{MqP_2}{r^4} - \frac{2}{3}\frac{J^2P_2}{r^4} + V_\mu , \\
    \psi_2 &=& \frac{qP_2}{r^3} + 3 \frac{MqP_2}{r^4} - 8\frac{J^2P_2}{r^4} \nonumber \\
           &+& \frac{1}{24}\frac{q^2}{r^6}\left(16 (P_2)^2 + 16 P_2 - 77 \right) + X_\mu ,  \nonumber\\
    \psi_3 &=& \frac{qP_2}{r^3} + \frac{5}{2} \frac{MqP_2}{r^4} - \frac{1}{2}\frac{J^2P_2}{r^4} \nonumber \\
           &+& \frac{1}{72}\frac{q^2}{r^6}\left(28 (P_2)^2 - 8 P_2 + 43 \right) + Y_\mu .  \nonumber
\end{eqnarray}

The electric charge $q_e$ and magnetic dipole $\mu_{d}$ contributions are given in the functions

\begin{eqnarray}
    V_\mu &=& \frac{1}{2}\frac{\mu_{d}^2 \cos^2 \theta}{r^4}  , \nonumber \\
    W_\mu &=& \frac{\mu_{d} q_e \sin^2 \theta}{r^2}   , \nonumber \\
    X_\mu &=& \frac{1}{3}  \frac{\mu_{d} ^2 (P_2 - 1) }{r^4}, \\
    Y_\mu &=& - \frac{1}{6}\frac{\mu_d^6 P_2}{r^4} .\nonumber
\end{eqnarray}

and the electromagnetic four-potential $A_\mu = (A_t, 0, 0, A_\phi)$ is

\begin{eqnarray}
    A_t &=& -\frac{q_e}{r} - \frac{J\mu_d \cos^2 \theta}{r^4} -\frac{qq_eP_2}{r^4}, \\
    A_\phi &=& \left(  \frac{\mu_d}{r} - \frac{3}{2} \frac{Jq_e}{r^2}  + \frac{3}{2} \frac{M\mu_d}{r^2} + \frac{1}{4} \frac{\mu_d q (P_2 +1) }{r^4}   \right) \sin^2 \theta . \nonumber
\end{eqnarray}

The equations of motion for test particles must be modified to include the electromagnetic contribution by means of the \emph{minimal coupling} and the \emph{super-Hamiltonian formalism} \cite{goldstein, gravitationmisnerthorne}. First, the generalized conjugate momentum associated with the test particle position $x^\mu$ takes the form $\pi_\mu = p_\mu + q_t A_\mu$, where $p_\mu$ is the \emph{classical four-momentum}, $q_t$ is the electric charge of the test particle, and $A_\mu$ is the electromagnetic potential produced by the source. Second, the relativistic energy--momentum relation is still conserved, $g^{\mu \nu} p_\mu p_\nu = -\mu^2$. Third, because the metric potentials and the electromagnetic potential depend on $r$ and $\theta$, but not of $t$ and $\phi$, there are two conserved momenta, $\pi_t = -E = p_t + q_t A_t$ and $\pi_\phi = L_z = p_\phi + q_t A_\phi$. Next, the super-Hamiltonian in terms of the particle rest mass $\mu$ and the momenta is

\begin{eqnarray}
    \mathcal{H} &=& \frac{1}{2} \frac{g^{\mu \nu} p_\mu p_\nu }{\mu} \nonumber \\
                &=& \frac{1}{2} \frac{g^{\mu \nu} (\pi_\mu - q_{t} A_\mu) (\pi_\nu - q_{t} A_\nu)}{\mu} \nonumber  \\
                &=& \frac{-\mu}{2} \nonumber \\
                &=& -\frac {(L_z - q_t A_{\phi} )^2 V} {2 \mu \rho^2 }   \nonumber  \\
                &+& \frac {  ( L_z - q_t A_{\phi} ) ( q_t A_t + E) W} {\mu \rho^2 }   \nonumber  \\
                &+& \frac {(\pi_r - q_t A_r ) {}^2} {2 \mu X } +  \frac {(\pi_\theta -  q_t A_{\theta} ) {}^2} {2 \mu Y }     \nonumber   \\
                &+&  \frac { ( q_t A_t  +  E)^2 Z} {2 \mu \rho^2 }    
\end{eqnarray}

where $\rho^2 = VZ + W^2$

Note that $A_r = A_\theta = 0$; hence, $\pi_r = p_r$ and $\pi_\theta = p_\theta$. Finally, the canonical equations of Hamilton are given by

\begin{eqnarray}
    \dot{r} &=& \frac { \pi_r - q_t A_r } {\mu X }  = \frac { p_r - q_t A_r } {\mu X } ,  \nonumber \\
    \dot{\theta} &=& \frac { \pi_\theta - q_t A_\theta  } {\mu Y} = \frac { p_\theta - q_t A_\theta  } {\mu Y}, \nonumber \\
    \dot{\phi} &=& \frac { (L_z -  q_t A_\phi) V -  (q_t A_t + E) W} {\mu \rho^2},   \nonumber \\
    \dot{t}  &=& \frac {(L_z - q_t A_\phi ) W } {\mu \rho^2} + \frac {(q_t A_t  + E) Z} {\mu \rho^2}. 
\end{eqnarray}

and

\begin{eqnarray}
\dot{\pi_r} &=& \dot{p_r} = -\frac {1} {2\mu} \left[   \frac {2 q_t (q_t A_t  + E) W \partial_r A_\phi  } {\rho^2 } \right. \nonumber \\
            &-& \frac {2 q_t  ( L_z - q_t A_\phi  ) V \partial_r A_\phi } {\rho^2 } - \frac {2 q_t (L_z - q_t A_\phi ) W  \partial_r A_t } { \rho^2 } \nonumber \\
            &+& \frac {2( L_z - q_t A_\phi ) ( q_t A_t  + E) W \partial_ r (\rho^2)} {\rho^4} \nonumber \\
            &-& \frac {2 (L_z - q_t A_\phi ) ( q_t A_t  + E ) \partial_r W } {\rho^2 }  \nonumber \\
            &-& \frac {(L_z - q_t A_\phi )^2 V \partial_r (\rho^2) } {\rho^4}  \nonumber \\
            &+& \frac {(L_z - q_t A_\phi )^2 \partial_r V } { \rho^2  } - \frac {2 q_t  ( \pi_r - q_t A_r )\partial_r A_r } {X}   \nonumber \\
            &-& \frac { ( \pi_r - q_t A_r )^2 \partial_r X } {X^2} - \frac {2 q_t ( q_t A_t  + E ) Z \partial_r A_t } { \rho^2 }   \nonumber \\
            &+& \frac { ( q_t A_t  + E)^2 Z \partial_r (\rho^2)} {\rho^4} - \frac {( q_t A_t  + E )^2 \partial_r Z } {\rho^2 }  \nonumber \\
            &-& \frac {2 q_t (\pi_\theta - q_t A_\theta ) \partial_r A_\theta } {Y } - \left. \frac { (\pi_\theta - q_t A_\theta )^2 \partial_r Y} {Y^2} \right],  \nonumber\\
\dot{\pi_\theta} &=& \dot{p_\theta} = -\frac {1} {2\mu } \left[\frac {2 q_t (q_t A_t + E) W \partial_\theta A_\phi} {\rho^2}  \right. \nonumber \\ 
            &-& \frac {2 q_t (L_z - q_t A_\phi) V \partial_\theta A_\phi} {\rho^2} - \frac {2 q_t (L_z - q_t A_\phi)W \partial_\theta A_t} {\rho^2} \nonumber \\ 
            &+& \frac {2 (L_z - q_t A_\phi ) (q_t A_t + E) W \partial_\theta (\rho^2)} {\rho^4} \nonumber \\
            &-& \frac {2 (L_z - q_t A_\phi) (q_t A_t + E) \partial_\theta W} {\rho^2} \nonumber \\
            &-& \frac {(L_z - q_t A_\phi)^2V \partial_\theta(\rho^2)} {\rho^4} \nonumber \\
            &+& \frac {(L_z - q_t A_\phi)^2 \partial_\theta V} {\rho^2} - \frac{2 q_t ( \pi_r - q_t A_r) \partial_\theta A_r} {X}  \nonumber \\
            &-& \frac {(\pi_r - q_t A_r)^2 \partial_\theta X} {X^2} - \frac {2 q_t (q_t A_t + E)Z \partial_\theta A_t } { \rho^2} \nonumber \\
            &+& \frac { ( q_t A_t + E)^2 Z  \partial_\theta (\rho^2) } {  \rho^4  } - \frac { (q_t A_t + E)^2  \partial_\theta Z } { \rho^2}  \nonumber \\
            &-& \frac {2 q_t ( \pi_\theta - q_t A_\theta) \partial_\theta A_\theta} {Y} - \left. \frac {(\pi_\theta - q_t A_\theta )^2\partial_\theta Y } {Y^2} \right], \nonumber \\
\dot{\pi_\phi} &=& 0,  \nonumber \\
\dot{\pi_t} &=& 0.  
\end{eqnarray}

In the equations of motion, the upper dot indicates differentiation with respect to the affine parameter. Real compact objects are expected to have a total electric charge $q_e \rightarrow 0$, as well as a vanishing test particle charge $q_t = 0$; therefore, we take them to be zero for the remainder of this work. However, there are several mechanisms by which compact objects can have a non-zero magnetic dipole moment. The Zeeman effect shows that even main-sequence stars may possess a magnetic field. Moreover, magnetars and pulsars exhibit the most powerful magnetic fields in the Universe \cite{LeBlancAstro}. 

In the next section, the magnetic contribution for both the KL and HT metrics will be considered by using their extended versions.

\section{\label{sec:DynA}Comparison of the dynamical systems}

The configuration space of a dynamical system consists of a $N$ manifold $M^N$ given by the $x^\mu$ local generalized coordinates. The Langrangian constructed with these coordinates is a real-valued function over the tangent bundle to $M$, while the Hamiltonian is a real-valued function on the phase space also known as the cotangent bundle to $M$. The phase space is an example of a symplectic manifold because of the existence of a non-degenerate 2-form $\omega^2 = dp_\mu \wedge dx^\mu$ \cite{Frankel}.  

Poincar\'e sections contain all the dynamical information of Hamiltonian systems. By taking the phase space of a system, the Poincar\'e section may be constructed. In the cases dealt with in this work, the configuration space has dimension $N = 4$, and the phase space has dimension $2N = 8$. A section of the phase space is taken in the equatorial plane for all $\phi$ angles and for positive $\theta$ angles. Poincar\'e sections transform a problem of orbits into a two-dimensional fixed-point problem. A closed curve in a Poincar\'e section represents a stable geodesic in the configuration space. When the mass quadrupole moment $q \ne 0$, other structures such as higher-order islands, hyperbolic points, and chaotic zones appear. These structures are the result of the nonlinear effects of the dynamical system \cite{adrian}.

The Kolmogorov--Arnold--Moser theorem (KAM) states that if the bounded motion of an integrable Hamiltonian $\mathcal{H}_0$ is perturbed by a small $\Delta \mathcal{H}$ that renders the total Hamiltonian $\mathcal{H} = \mathcal{H}_0 + \Delta \mathcal{H}$ non-integrable, and if the perturbation $\Delta \mathcal{H}$ is small and the frequencies $\omega_i$ of $\mathcal{H}_0$ are incommensurate. Then, the motion remains bounded on an $N$-dimensional torus, except for a negligible set of initial conditions resulting in a meandering path over the energy surface \cite{adrian, goldstein, contopoulosorderandchaos}.

A KAM torus in a symplectic dynamical system exhibits quasiperiodic motion and is diffeomorphic to an $n$-torus, where $n = N/2$; in our case, $n = 4$. KAM theory also asserts that there exists a large set of invariant tori with quasiperiodic motion that persist under small perturbations of the Hamiltonian \cite{lazutkinKAMTheory}.

The Poincar\'e--Birkhoff (PB) theorem states that, from the infinitely many periodic orbits on a resonant torus of the integrable system, only an even number survive in the perturbed system. Moreover, half of these surviving periodic orbits are stable, while the other half are unstable \cite{adrian, LukesZpoyVoorhees, contopoulosorderandchaos}.

The KAM and PB theorems indicate how the phase space evolves as $\lvert q \rvert$ increases. The KAM theorem states that most of the surviving orbits will still be confined to a torus, although they will be slightly deformed. The PB theorem indicates that the rupture of the torus gives rise to new types of structures visible in a Poincar\'e section \cite{adrian, lazutkinKAMTheory}.

\subsection{Spacetime without electromagnetic contribution}

To compare the dynamics of a test particle using different metrics, a set of parameters similar to those of \cite{misbah} has been chosen. The gravitational source mass is $M = 1.0$, the gravitational source spin parameter is $a = J/M = 0.1$, the orbital energy is $E = 0.95$, the orbital angular momentum is $L_z = 3.0$, and the test particle mass is $\mu = 1.0$. This is analogous to using units of $M$ to measure distances. The selection of $a < M$ aims to ensure a slowly rotating gravitational source, while still being sufficient to produce a frame-dragging effect similar to that predicted by the Kerr metric.  

The last parameter, the mass quadrupole moment $q$, which represents the perturbation from the Kerr metric, is different for the two metrics. However, Eq.~(\ref{E:rel_q}) relates both types of quadrupole moments. By keeping the parameters $M$, $a$, $E$, $L_z$, and $m$ fixed, and by taking $q_{KL}$ as the control parameter (with $q_{HT}$ calculated using Eq.~(\ref{E:rel_q})), it is possible to make a direct comparison of the dynamical systems. In this first section, only the classical metrics (KL, HTlog, and appHT) without electromagnetic components are considered. The magnetic dipole moment of the compact object will be added in the next section.


First, it is useful to check the \emph{unperturbed} case, when $q_{KL} = 0$. In this case, the KL metric reduces to the standard Kerr metric and, as stated before, it does not show structures outside the main island of stability. On the other hand, neither HT metrics reduce exactly to the Kerr metric (the HT metrics reduce to the Lense--Thirring metric), and both HT metrics have $q_{HT} = 0.01$, which represents a small perturbation; consequently, the phase space exhibits small secondary structures.

\begin{figure}[h]
    \centering
    \includegraphics[width=0.9\linewidth]{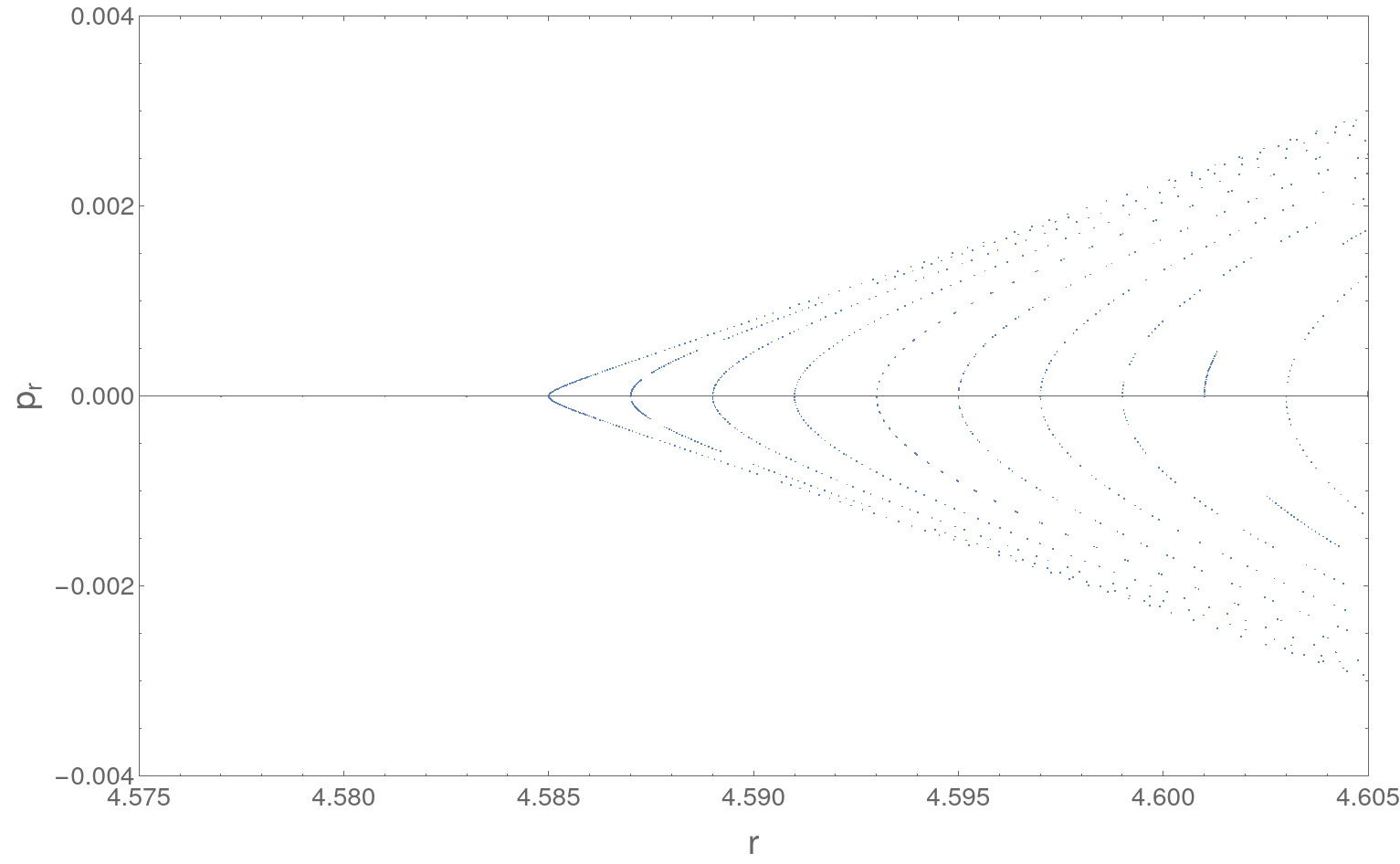}
    \caption{Poincar\'e section using KL metric for $\theta = \pi/2$ and $p_\theta \geq 0$ and parameters $M = 1.0$, $a = 0.1$, $E = 0.95$, $L_z = 3.0$ and $q_{KL} = 0$, there is no structure beside the main island of stability.}
    \label{fig:poinKLq0}
\end{figure}

\begin{figure}[h]
    \centering
    \includegraphics[width=0.9\linewidth]{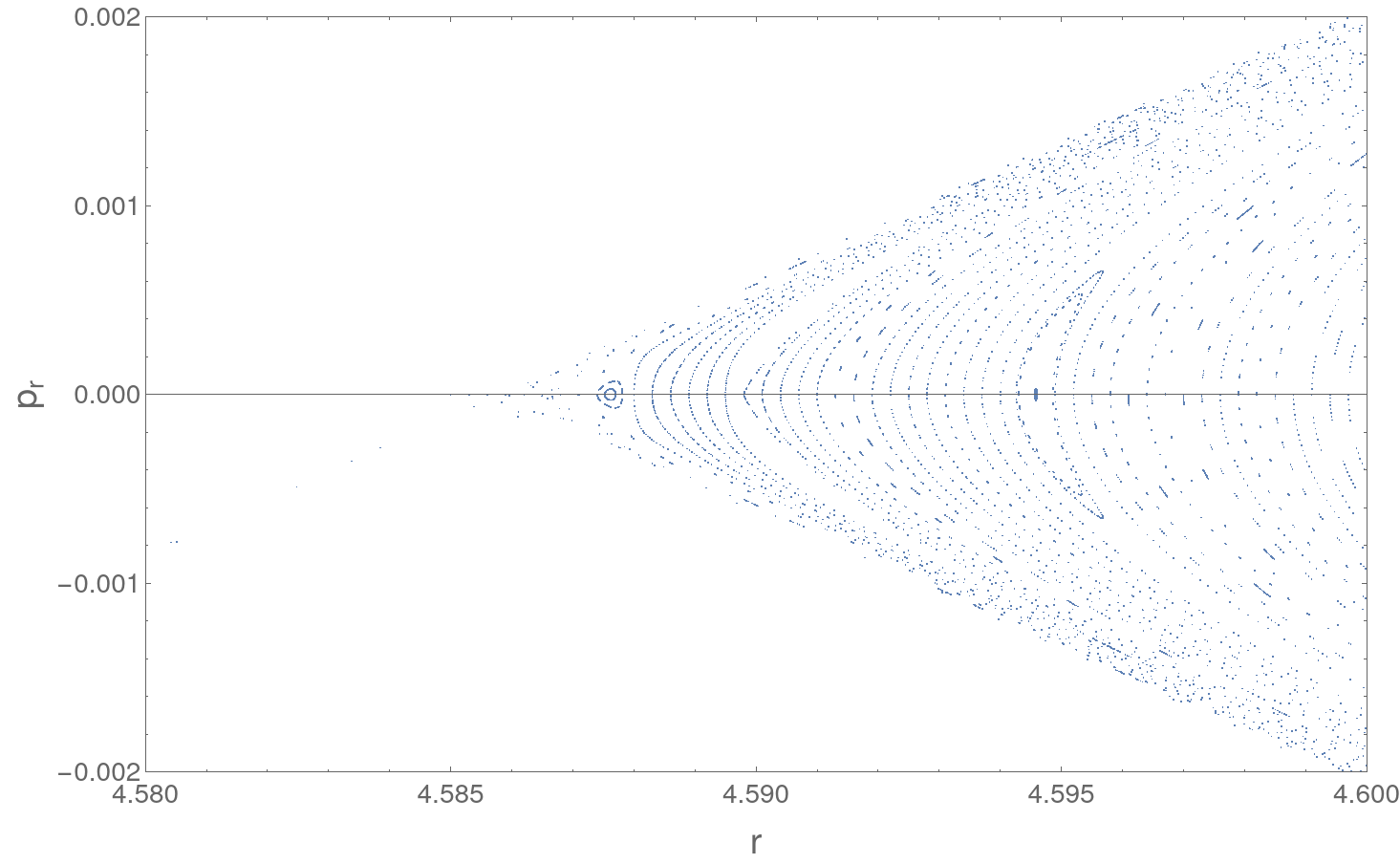}
    \caption{Poincar\'e section using HTlog metric for $\theta = \pi/2$ and $p_\theta \geq 0$ and parameters $M = 1.0$, $a = 0.1$, $E = 0.95$, $L_z = 3.0$ and $q_{KL} = 0$, there is a small structure on the tip nearest to the compact object (left) and another resonance near $r = 4.595$.}
    \label{fig:poinHTq0}
\end{figure}

\begin{figure}[h]
    \centering
    \includegraphics[width=0.9\linewidth]{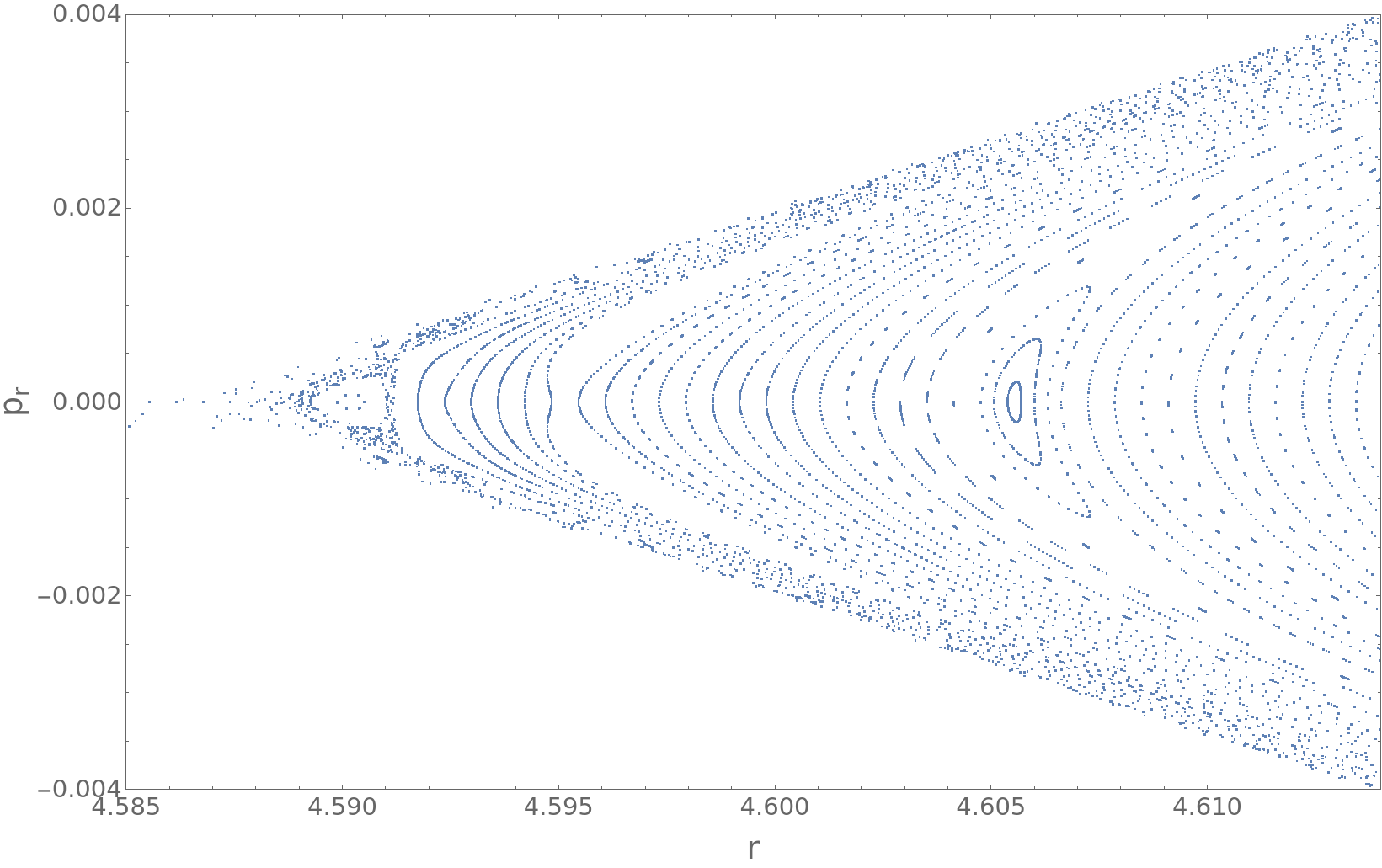}
    \caption{Poincar\'e section using appHT metric for $\theta = \pi/2$ and $p_\theta \geq 0$ and parameters $M = 1.0$, $a = 0.1$, $E = 0.95$, $L_z = 3.0$ and $q_{KL} = 0$. Again, there is a resonance on the tip nearest to the compact object surrounded by second order islands and chaos (left) and another resonance with center near $r = 4.606$.}
    \label{fig:appHTq0}
\end{figure}

Figure~\ref{fig:poinKLq0} represents the KL Poincar\'e section in the limit $q \rightarrow 0$, which corresponds to the Kerr case, an integrable system that exhibits only the main island of stability. In contrast, Fig.~\ref{fig:poinHTq0} shows that the HTlog metric presents a small region of incipient chaos and small first-order islands because $q_{KL} = 0$ implies $q_{HT} = 0.01$. Even with this small perturbation in the HTlog metric, some stable orbits are broken, as predicted by the Kolmogorov--Arnold--Moser and Poincar\'e--Birkhoff theorems \cite{adrian}, giving rise to narrow resonances and traces of chaos in the region $4.585 < r < 4.587$ and $p_r = 0$ (to the left of the island tip). The Kerr-like metric does not present this behavior in the same region; in fact, it remains completely stable. The appHT metric, for the same set of parameters, is shown in Fig.~\ref{fig:appHTq0}, exhibits a less stable tip, with a broader chaotic region, and the resonance centered at $r = 4.606$ is more prominent than its counterpart in the HTlog case. Nevertheless, the HT metrics are only approximations of the Kerr metric and are not expected to reproduce identical results, especially in regions where gravity is most intense. Globally, HTlog and appHT remain qualitatively similar to Kerr, as the main islands of stability in all three cases begin near $r = 4.585$.

As the intensity of the perturbation is increased, the differences between the metrics are enhanced for both prolate and oblate cases.

\begin{figure}[h]
    \centering
    \includegraphics[width=0.9\linewidth]{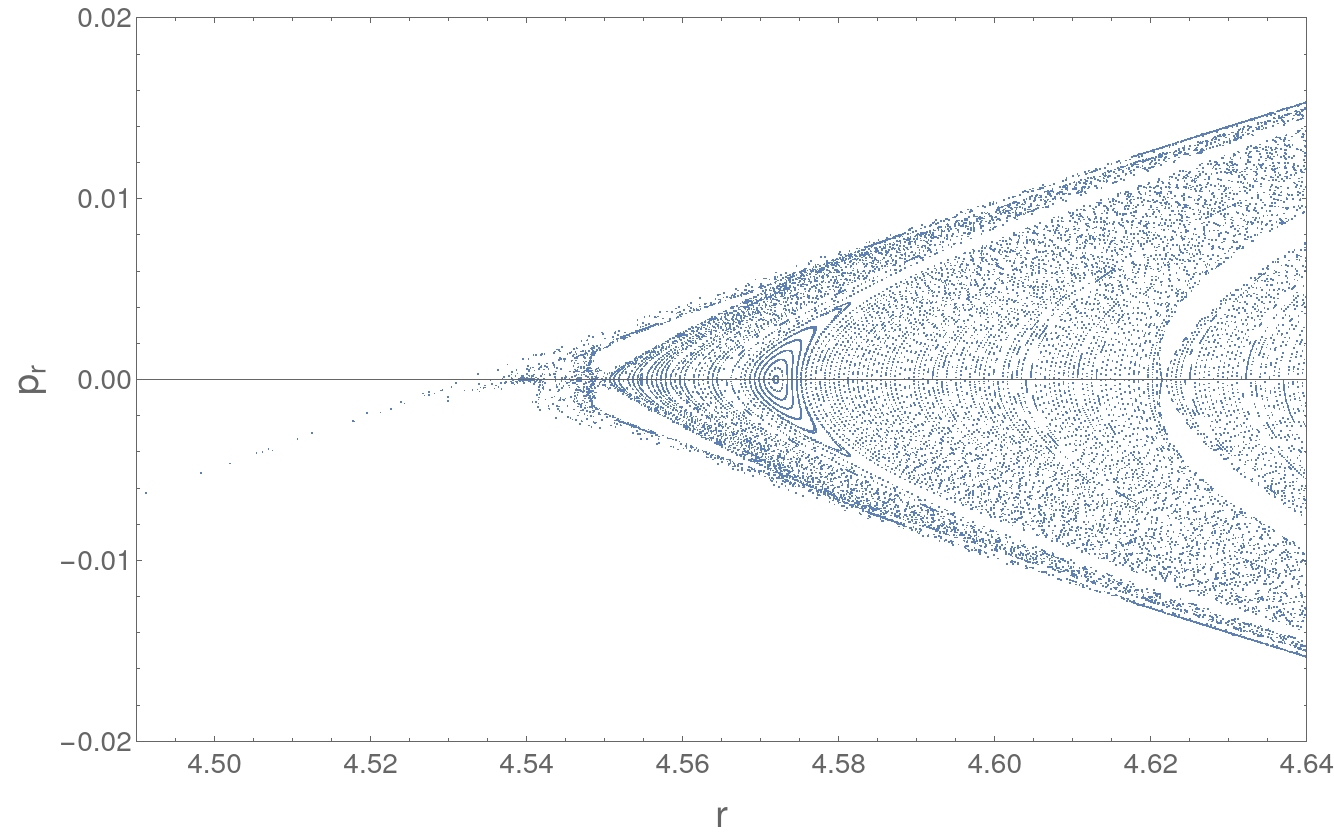}
    \caption{Poincar\'e section using KL metric for $\theta = \pi/2$ and $p_\theta \geq 0$ and parameters $M = 1.0$, $a = 0.1$, $E = 0.95$, $L_z = 3.0$ and $q_{KL} = 0.1$.}
    \label{fig:poinKLq01}
\end{figure}

\begin{figure}[h]
    \centering
    \includegraphics[width=0.9\linewidth]{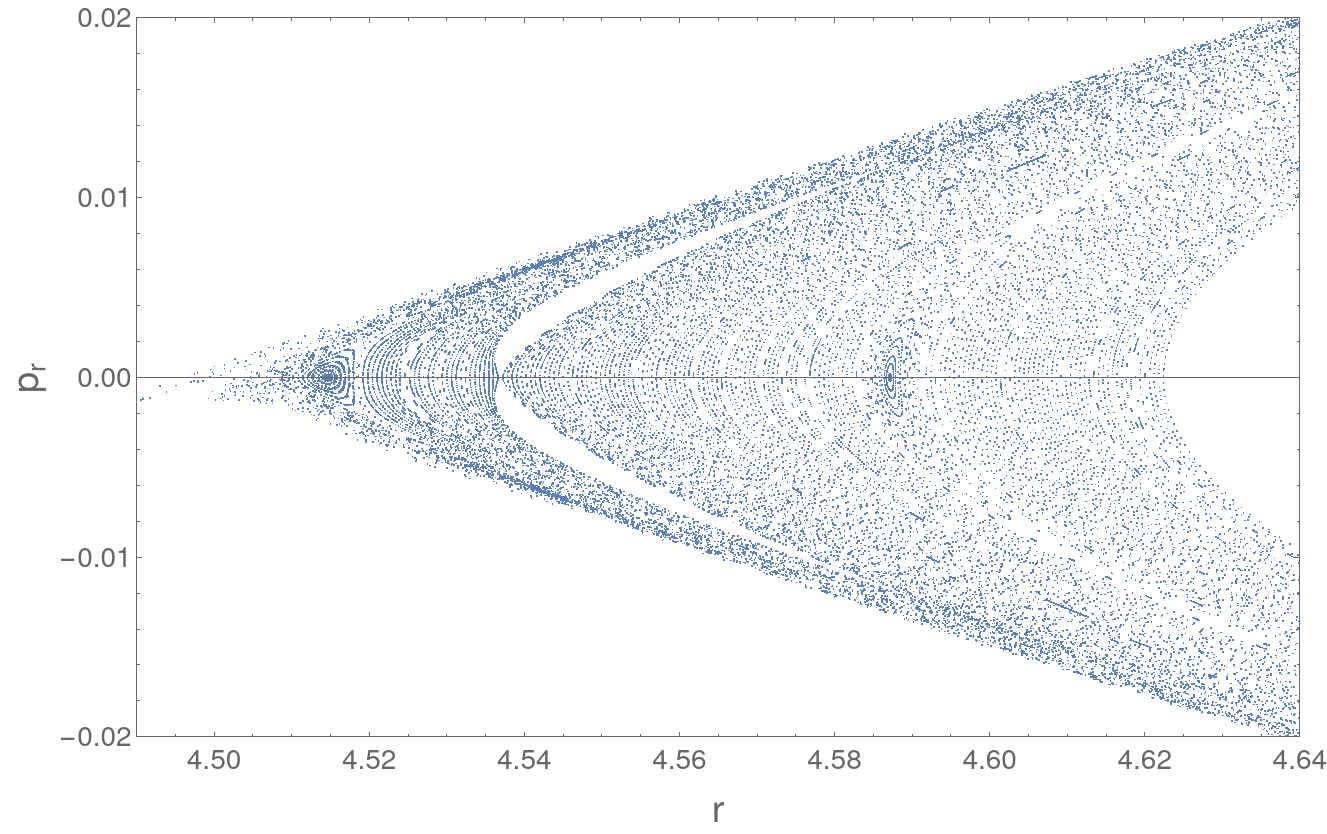}
    \caption{Poincar\'e section using HTlog metric for $\theta = \pi/2$ and $p_\theta \geq 0$ and parameters $M = 1.0$, $a = 0.1$, $E = 0.95$, $L_z = 3.0$ and $q_{KL} = 0.1$.}
    \label{fig:poinHTq01}
\end{figure}

\begin{figure}[h]
    \centering
    \includegraphics[width=0.9\linewidth]{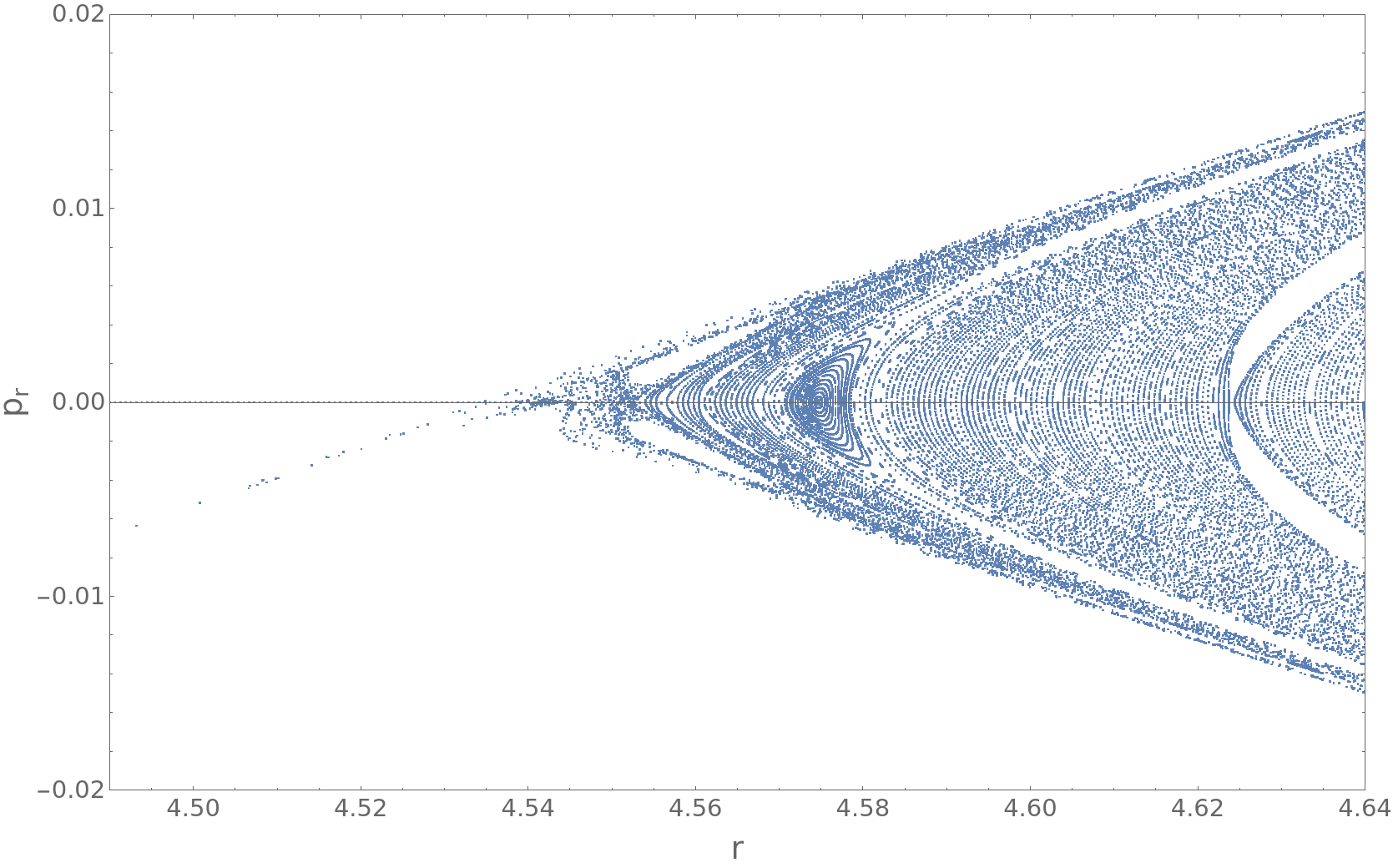}
    \caption{Poincar\'e section using appHT metric for $\theta = \pi/2$ and $p_\theta \geq 0$ and parameters $M = 1.0$, $a = 0.1$, $E = 0.95$, $L_z = 3.0$ and $q_{KL} = 0.1$.}
    \label{fig:appHTq01}
\end{figure}

\begin{figure}[h]
    \centering
    \includegraphics[width=0.9\linewidth]{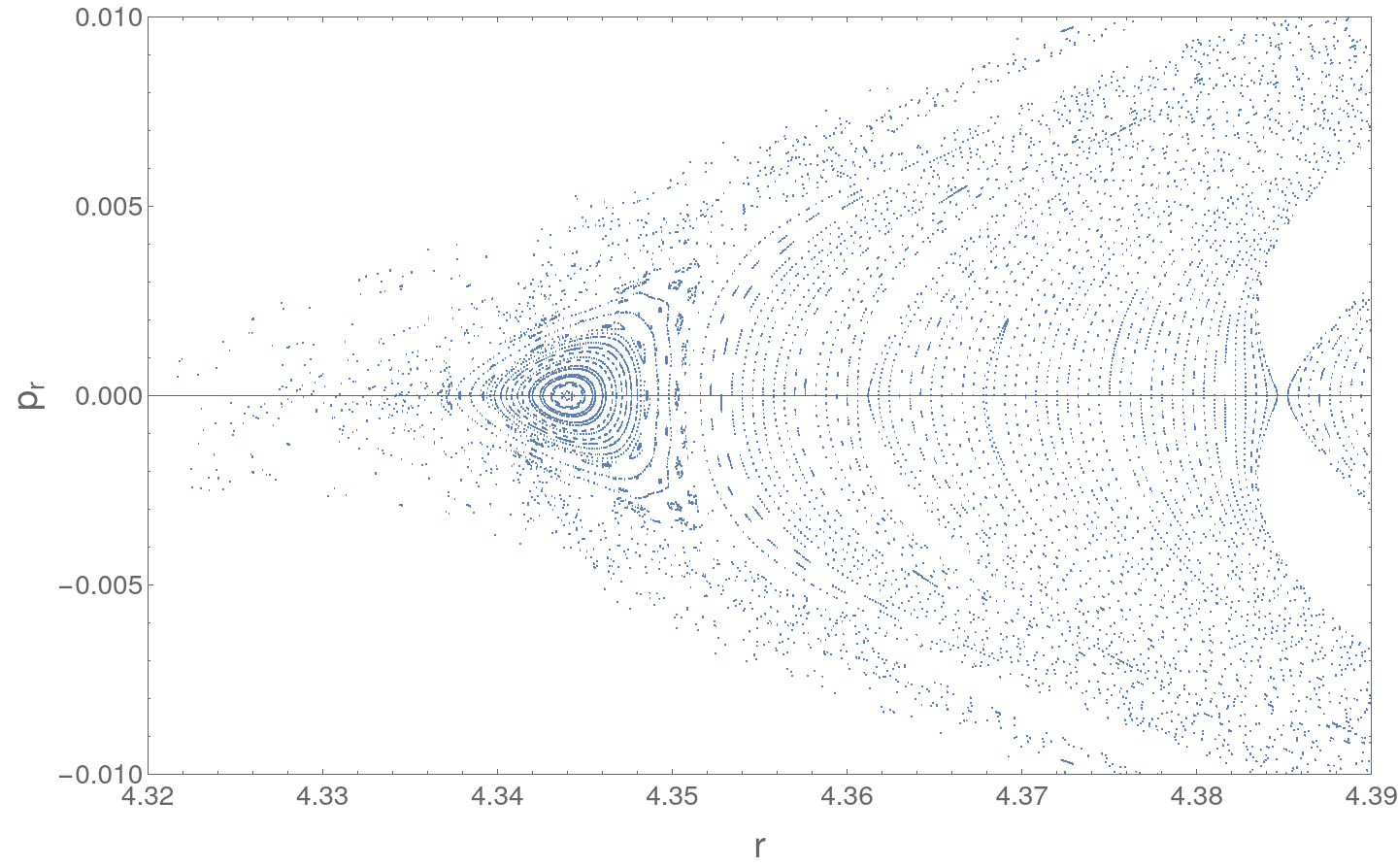}
    \caption{Poincar\'e section using KL metric for $\theta = \pi/2$ and $p_\theta \geq 0$ and parameters $M = 1.0$, $a = 0.1$, $E = 0.95$, $L_z = 3.0$ and $q_{KL} = 0.5$.}
    \label{fig:poinKLq05}
\end{figure}

\begin{figure}[h]
    \centering
    \includegraphics[width=0.9\linewidth]{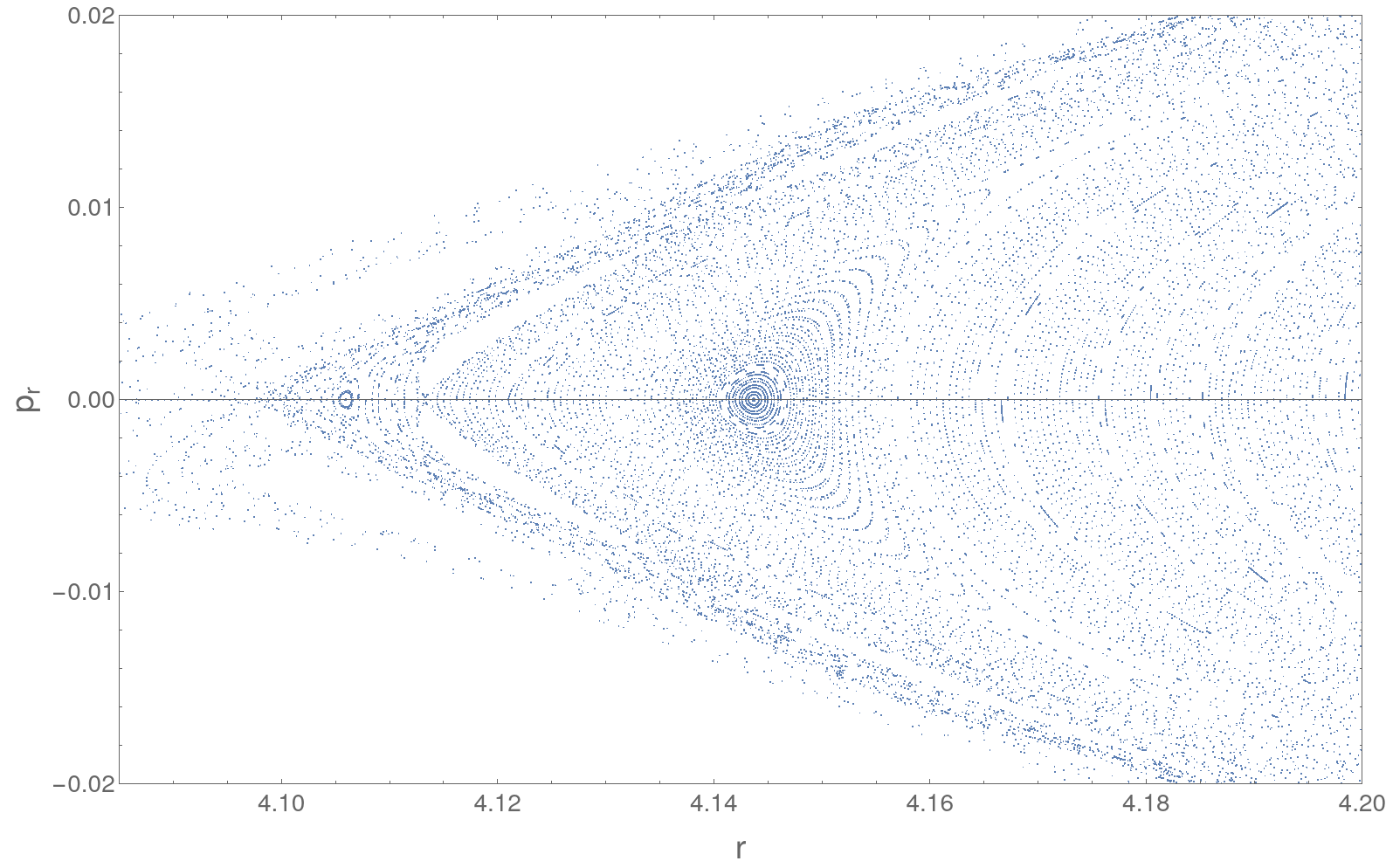}
    \caption{Poincar\'e section using HTlog metric for $\theta = \pi/2$ and $p_\theta \geq 0$ and parameters $M = 1.0$, $a = 0.1$, $E = 0.95$, $L_z = 3.0$ and $q_{KL} = 0.5$.}
    \label{fig:poinHTq05}
\end{figure}

\begin{figure}[h]
    \centering
    \includegraphics[width=0.9\linewidth]{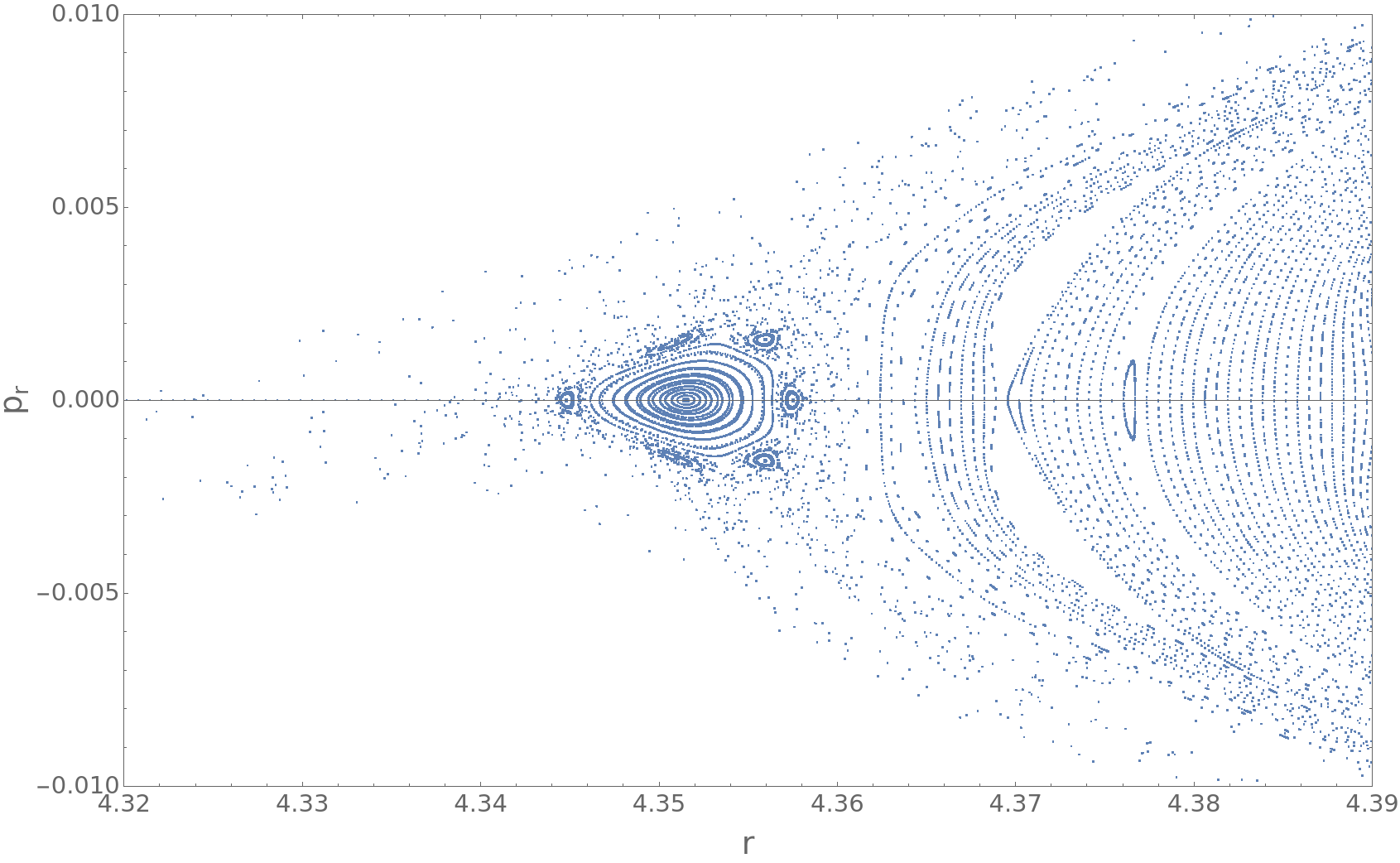}
    \caption{Poincar\'e section using appHT metric for $\theta = \pi/2$ and $p_\theta \geq 0$ and parameters $M = 1.0$, $a = 0.1$, $E = 0.95$, $L_z = 3.0$ and $q_{KL} = 0.5$.}
    \label{fig:appHTq05}
\end{figure}

For $q_{KL} = 0.1$, which implies $q_{HT} = -0.09$, the KL metric shown in Fig.~\ref{fig:poinKLq01} exhibits a rich variety of structures and chaotic regions around the tiny island located at $r = 4.54$, whereas the HTlog Poincar\'e section in Fig.~\ref{fig:poinHTq01} presents a more stable behavior. In this case, the HTlog metric spans the region $4.50 < r < 4.53$, where the KL metric does not. By contrast, the appHT section shown in Fig.~\ref{fig:appHTq01} matches the KL results very well. This behavior can be explained by the fact that the appHT metric lacks the complicated logarithmic factors, and its Taylor expansion agrees more closely with the Taylor expansion of the KL metric, whereas the HTlog metric shows a stronger deviation with respect to KL.

A similar pattern is observed when $q_{KL} = 0.5$ and $q_{HT} = -0.49$. In these cases, the gravitational source is a prolate object because $q_{KL} > 0$. Figures~\ref{fig:poinKLq05} and~\ref{fig:appHTq05} show the Poincar\'e sections for the KL and appHT metrics, respectively. Although both share similarities, such as their main islands of stability starting near $r = 4.335$, with islands at this location surrounded by satellites, clear differences also appear. First, the islands at this tip exhibit a divergent behavior. For appHT, this island is more sharply defined and displays six smaller second-order islands surrounded by a chaotic layer. On the other hand, the corresponding island for KL is surrounded by many second-order islands and is enclosed by narrow stable orbits, as well as higher-order islands and chaotic regions. Second, the location of the hyperbolic point differs: for KL it is located near $r = 4.385$, whereas for appHT it appears near $r = 4.37$. 

Figure~\ref{fig:poinHTq05} shows the Poincar\'e section for the HTlog metric. In this case, the main island of stability begins near $r = 4.10$ and exhibits structures that differ from the previous sections, even though the same parameters and coordinates are used. The HTlog metric is better suited for very slowly rotating and slightly deformed gravitational sources, whereas appHT appears to have a wider range of applicability for larger values of the mass quadrupole moment.

\begin{figure}[h]
    \centering
    \includegraphics[width=0.9\linewidth]{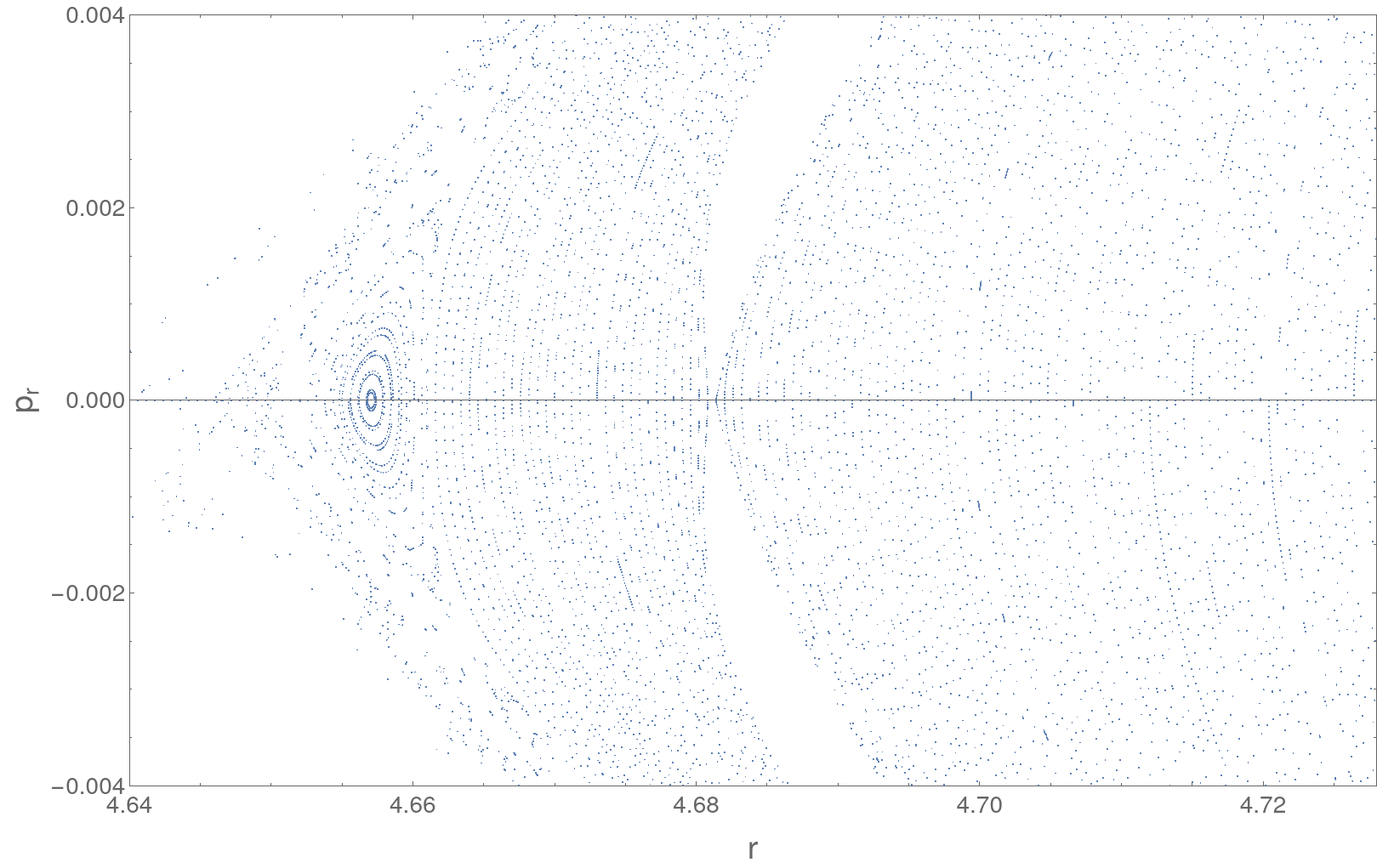}
    \caption{Poincar\'e section using KL metric for $\theta = \pi/2$ and $p_\theta \geq 0$ and parameters $M = 1.0$, $a = 0.1$, $E = 0.95$, $L_z = 3.0$ and $q_{KL} = -0.1$.}
    \label{fig:poinKLq-01}
\end{figure}

\begin{figure}[h]
    \centering
    \includegraphics[width=0.9\linewidth]{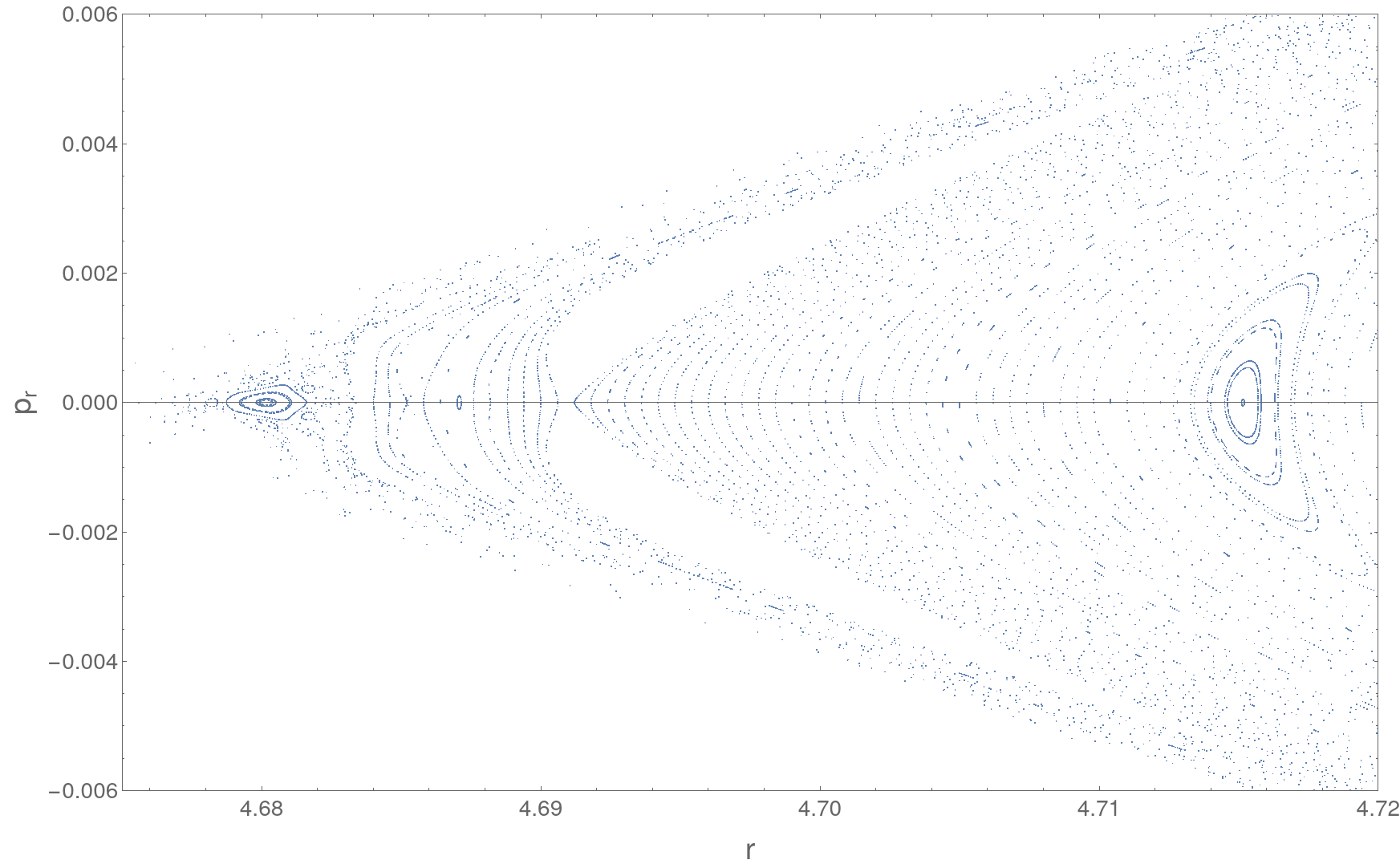}
    \caption{Poincar\'e section using HTlog metric for $\theta = \pi/2$ and $p_\theta \geq 0$ and parameters $M = 1.0$, $a = 0.1$, $E = 0.95$, $L_z = 3.0$ and $q_{KL} = -0.1$.}
    \label{fig:poinHTq-01}
\end{figure}

\begin{figure}[h]
    \centering
    \includegraphics[width=0.9\linewidth]{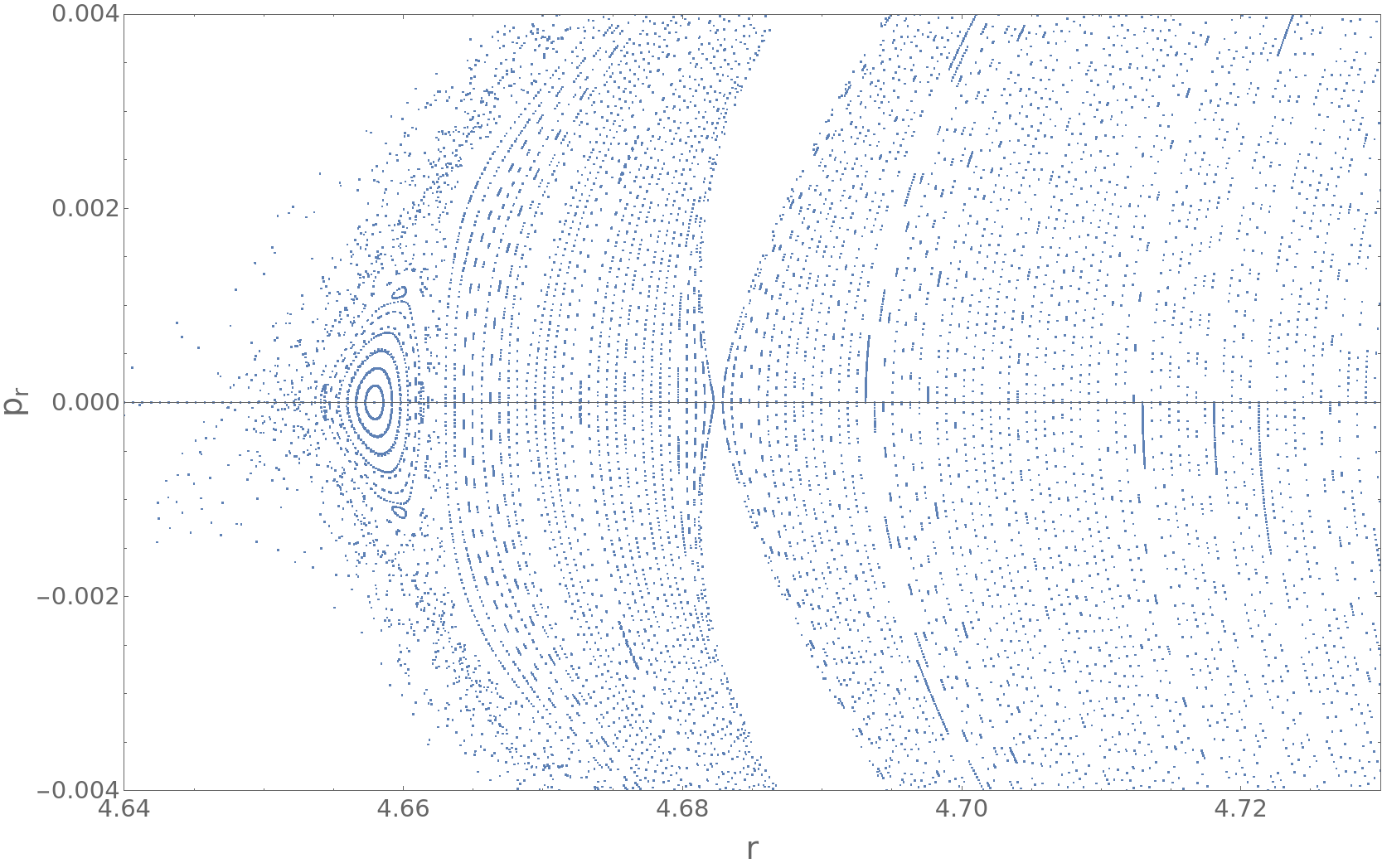}
    \caption{Poincar\'e section using appHT metric for $\theta = \pi/2$ and $p_\theta \geq 0$ and parameters $M = 1.0$, $a = 0.1$, $E = 0.95$, $L_z = 3.0$ and $q_{KL} = -0.1$.}
    \label{fig:appHTq-01}
\end{figure}

\begin{figure}[h]
    \centering
    \includegraphics[width=0.9\linewidth]{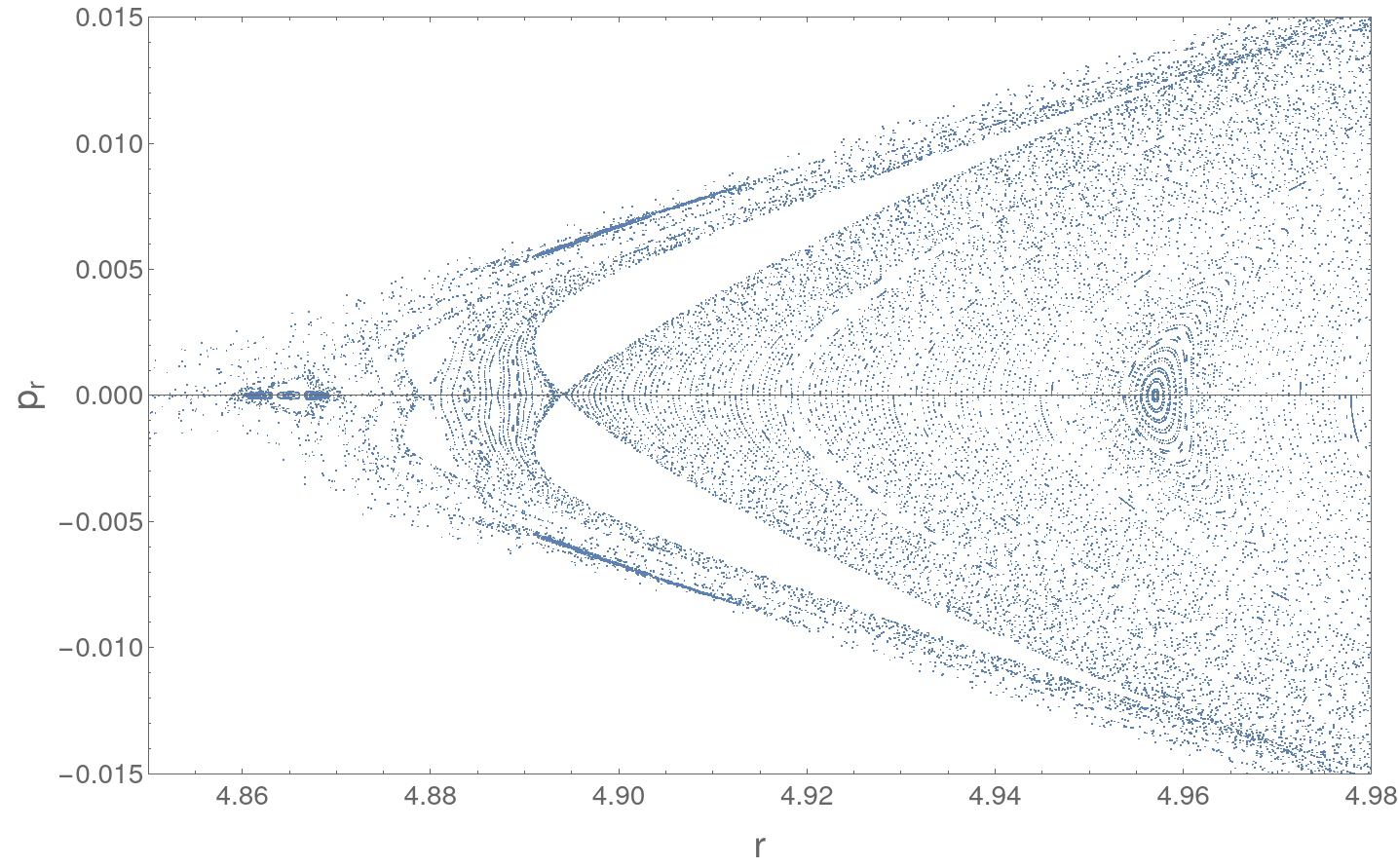}
    \caption{Poincar\'e section using KL metric for $\theta = \pi/2$ and $p_\theta \geq 0$ and parameters $M = 1.0$, $a = 0.1$, $E = 0.95$, $L_z = 3.0$ and $q_{KL} = -0.5$.}
    \label{fig:poinKLq-05}
\end{figure}

\begin{figure}[h]
    \centering
    \includegraphics[width=0.9\linewidth]{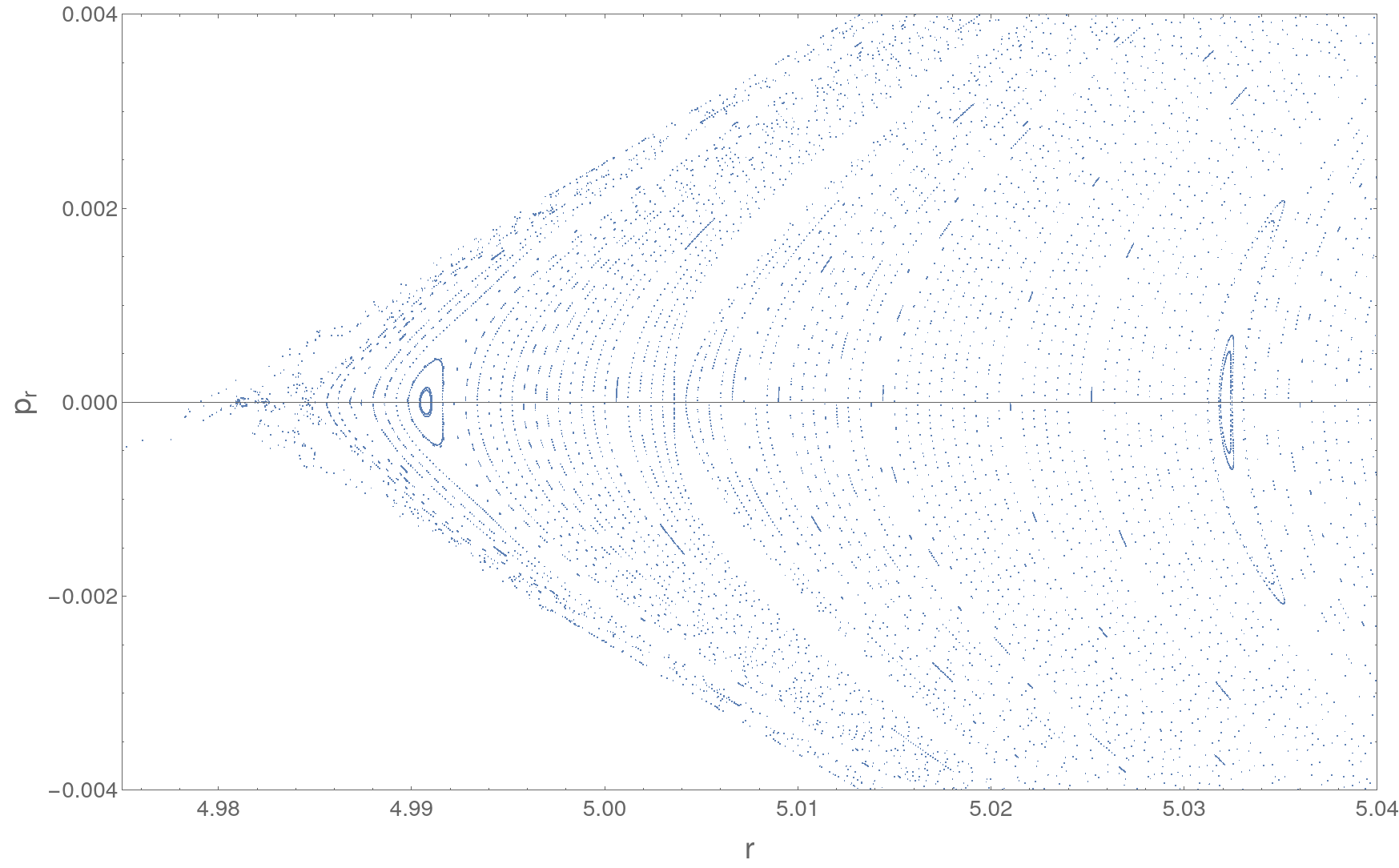}
    \caption{Poincar\'e section using HTlog metric for $\theta = \pi/2$ and $p_\theta \geq 0$ and parameters $M = 1.0$, $a = 0.1$, $E = 0.95$, $L_z = 3.0$ and $q_{KL} = -0.5$.}
    \label{fig:poinHTq-05}
\end{figure}

\begin{figure}[h]
    \centering
    \includegraphics[width=0.9\linewidth]{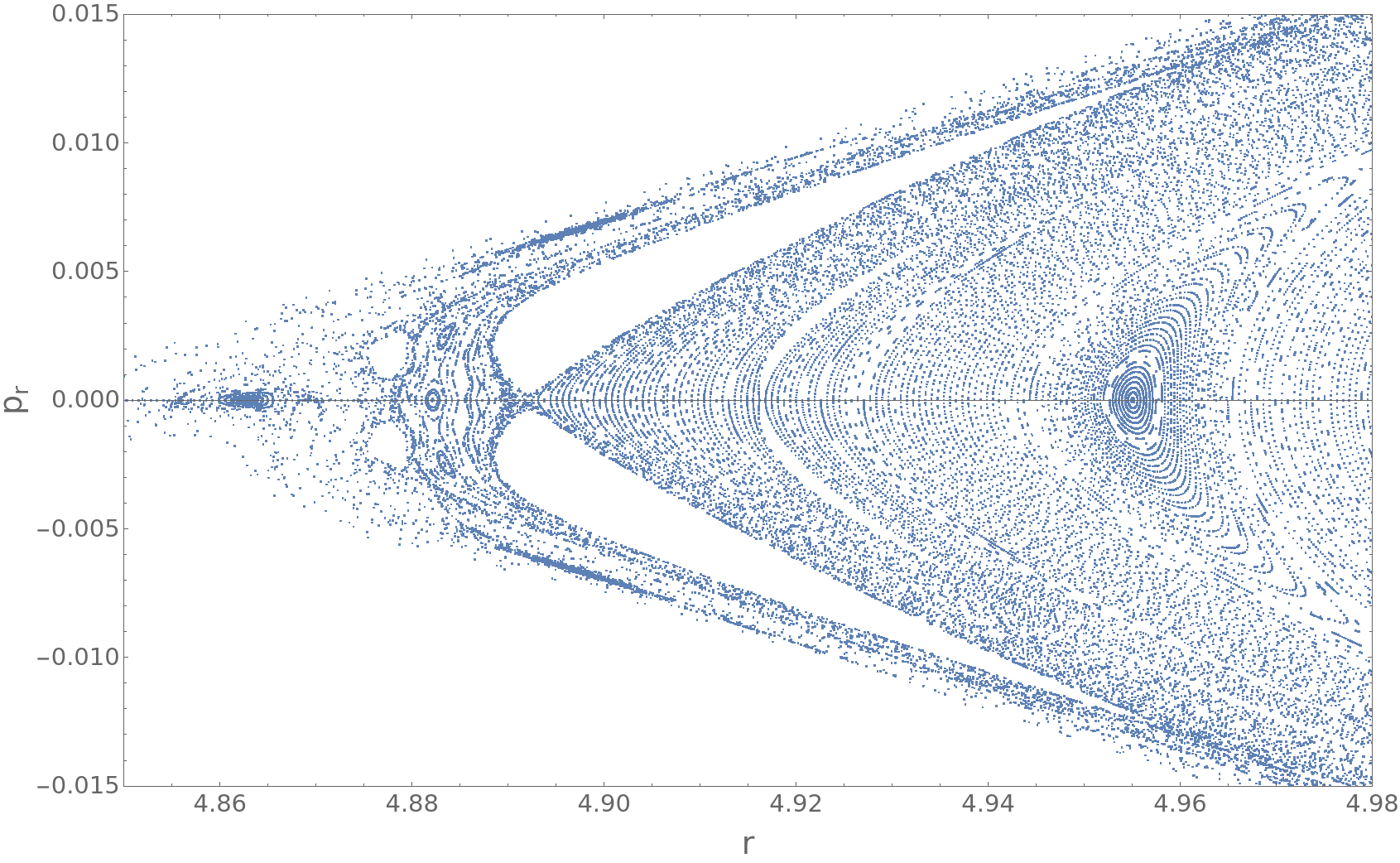}
    \caption{Poincar\'e section using appHT metric for $\theta = \pi/2$ and $p_\theta \geq 0$ and parameters $M = 1.0$, $a = 0.1$, $E = 0.95$, $L_z = 3.0$ and $q_{KL} = -0.5$.}
    \label{fig:appHTq-05}
\end{figure}

For the case of oblate objects with $q_{KL} = -0.1$ and $q_{HT} = 0.11$, the inverse behavior is observed. The KL metric (figure~\ref{fig:poinKLq-01}) and the appHT metric (figure~\ref{fig:appHTq-01}) both produce Poincar\'e sections that are more stable than their HTlog counterpart (figure~\ref{fig:poinHTq-01}). The KL and appHT sections show very few structures near the tip at $r = 4.65$, whereas HTlog presents several higher-order islands, hyperbolic points, and more pronounced traces of chaotic geodesics. Additionally, the tip of the main island of stability for HTlog is slightly shifted outward from the gravitational source, near $r = 4.66$, and the section exhibits several hyperbolic points.

For the case of $q_{KL} = -0.5$ and $q_{HT} = 0.51$, the KL section of figure~\ref{fig:poinKLq-05} and the appHT section of figure~\ref{fig:appHTq-05} present several structures that are similar to each other. Their corresponding structures are only slightly shifted and deformed; for example, in the KL case the island at the left tip ($r = 4.86$) is surrounded by two larger second-order islands, while the appHT first-order island at this position has two smaller satellites. Other corresponding structures are more visually prominent in appHT than in KL. On the other hand, the HTlog section of figure~\ref{fig:poinHTq-05} shows a divergent behavior with respect to the KL and appHT cases. The phase space for HTlog exhibits fewer structures because more tori are completely destroyed by the effect of the mass quadrupole moment. The main island of stability starts near $r = 4.98$, with a tiny first-order island surrounded by traces of chaos. In addition, the remaining structures do not match those of the other metrics.


\subsection{Effects of the Magnetic Dipole Moment}

In this section, an additional perturbation is considered: the magnetic dipole moment $\mu_{d}$ of the gravitational source. Electromagnetic fields may curve spacetime. For these cases, the Kerr--Newman--like metric (KLdip) of equation~(\ref{E:KLdip}) and an extension of the Hartle--Thorne metric with magnetic dipole moment (HTdip) of equation~(\ref{E:HTdip}) are considered, using the previous parameters but adding $\mu_{d}$. For every Poincar\'e section in this part, only uncharged gravitational sources ($q_e = 0$) and uncharged test particles ($q_t = 0$) are considered.

\begin{figure}[h]
    \centering
    \includegraphics[width=0.9\linewidth]{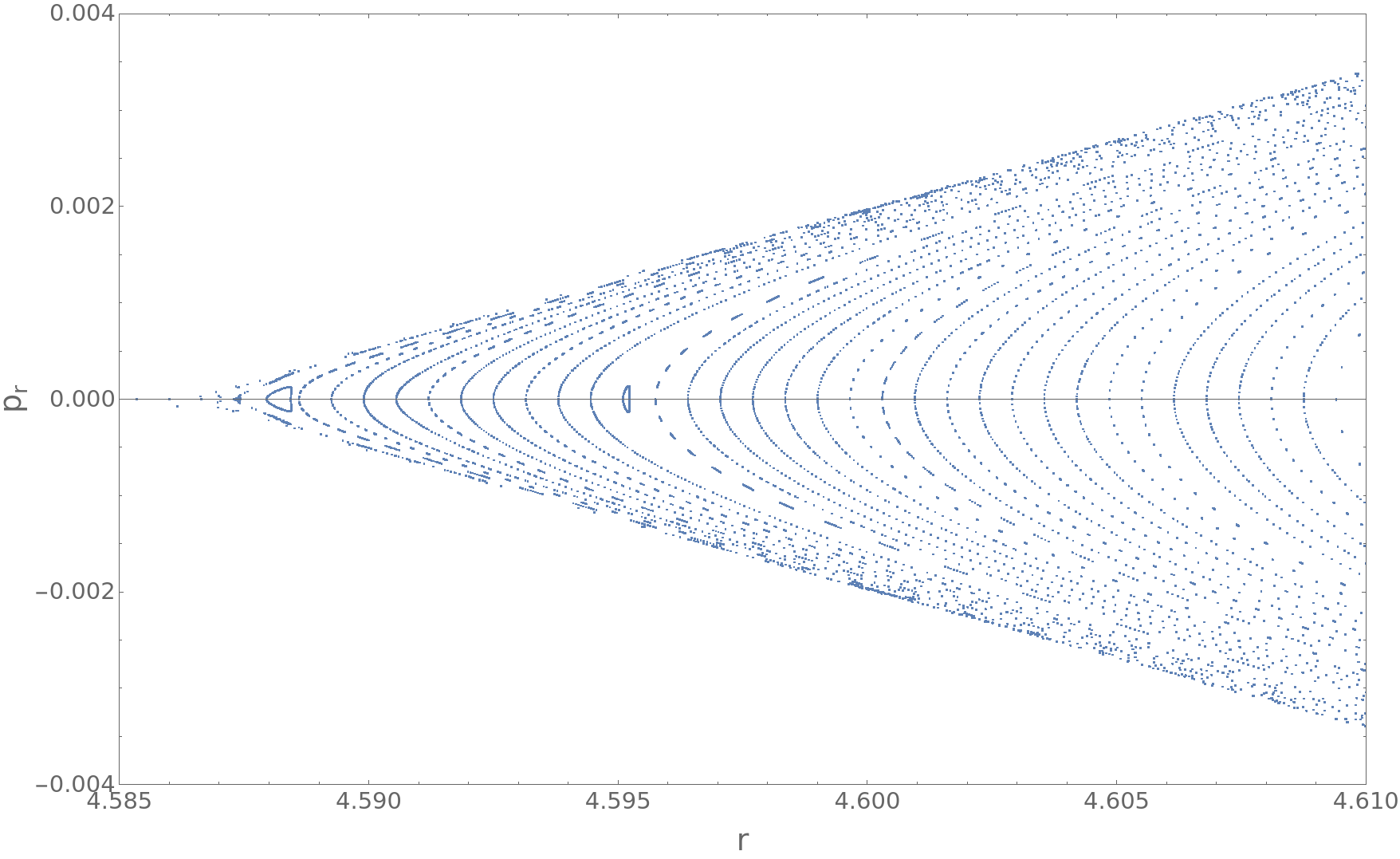}
    \caption{Poincar\'e section using KLdip metric for $\theta = \pi/2$ and $p_\theta \geq 0$ and parameters $M = 1.0$, $a = 0.1$, $E = 0.95$, $L_z = 3.0$, $\mu_{d} = 0.2$ and $q_{KL} = 0$.}
    \label{fig:KLdipmu02q0}
\end{figure}

\begin{figure}[h]
    \centering
    \includegraphics[width=0.9\linewidth]{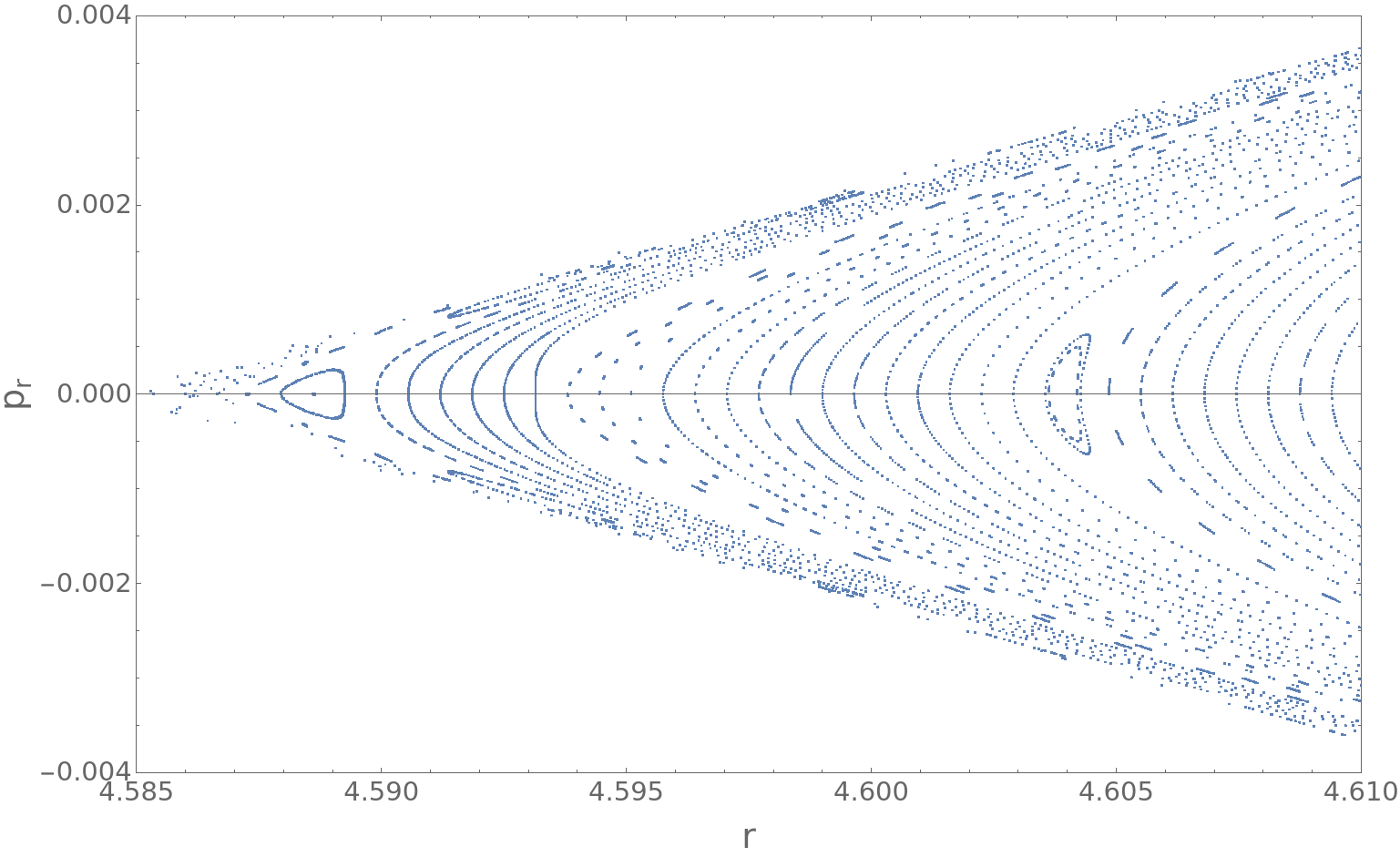}
    \caption{Poincar\'e section using HTdip metric for $\theta = \pi/2$ and $p_\theta \geq 0$ and parameters $M = 1.0$, $a = 0.1$, $E = 0.95$, $L_z = 3.0$, $\mu_{d} = 0.2$, $q_{KL} = 0$.}
    \label{fig:HTdipmu02q0}
\end{figure}

Figures~\ref{fig:KLdipmu02q0} and \ref{fig:HTdipmu02q0} show the effect of the magnetic dipole moment $\mu_d = 0.2$ on the phase space for KLdip and HTdip, respectively, which clearly exhibit structures and chaos, as expected from a small perturbation of an integrable Hamiltonian. Both sections were constructed for a spherical compact object with $q_{KL} = 0$. As shown in these figures, there are similarities in the phase space; for example, their main islands of stability start near $r = 4.587$, with a tiny resonance followed by a larger one centered near $r = 4.588$. Nevertheless, this island is larger in HTdip than in KLdip. Moreover, HTdip shows additional structures, such as a resonance at $r = 4.604$, while KLdip does not.

\begin{figure}[h]
    \centering
    \includegraphics[width=0.9\linewidth]{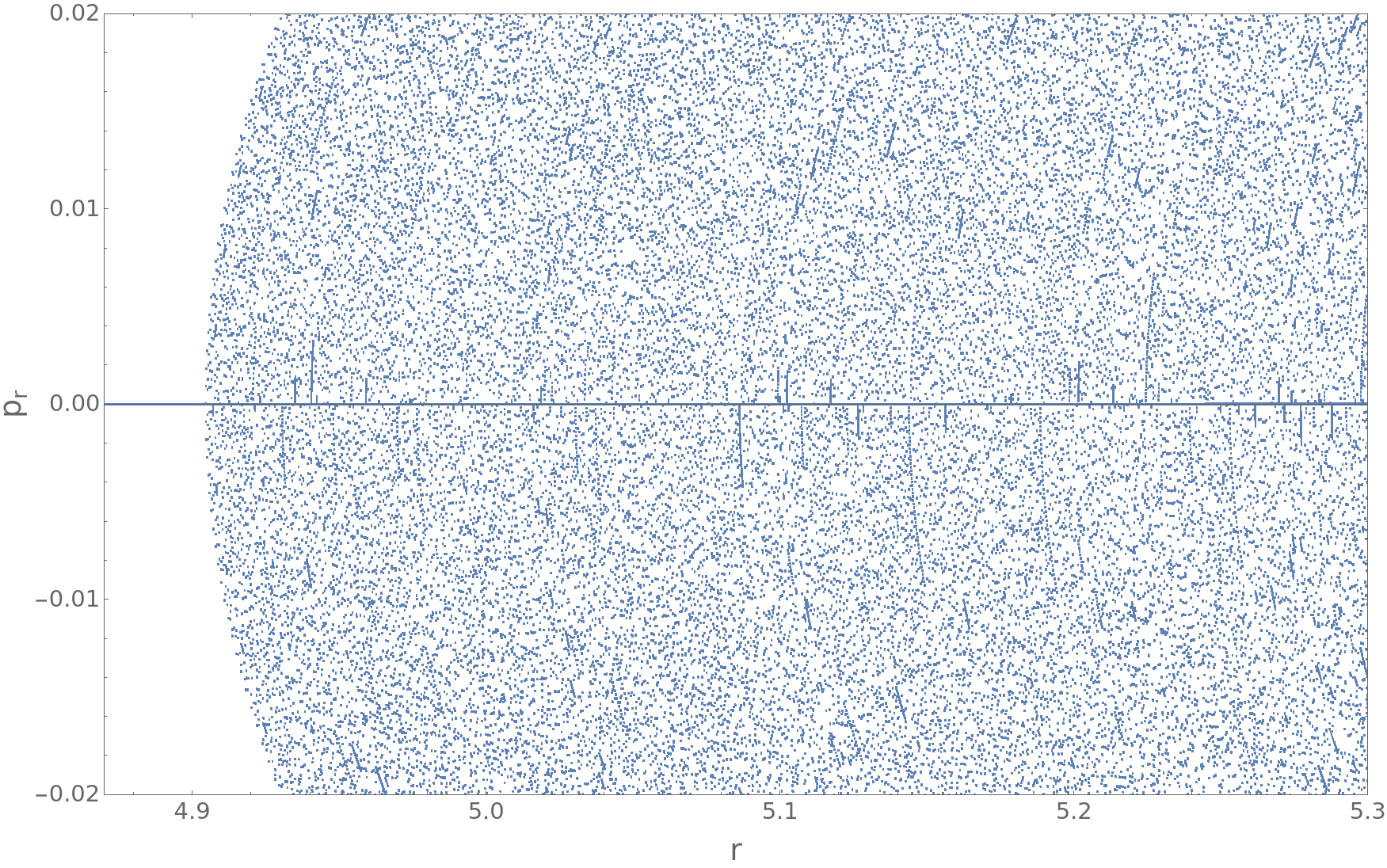}
    \caption{Poincar\'e section using KLdip metric for $\theta = \pi/2$ and $p_\theta \geq 0$ and parameters $M = 1.0$, $a = 0.1$, $E = 0.95$, $L_z = 3.0$, $\mu_{d} = 0.2$ and $q_{KL} = 0.5$.}
    \label{fig:KLdipmu02q05}
\end{figure}

\begin{figure}[h]
    \centering
    \includegraphics[width=0.9\linewidth]{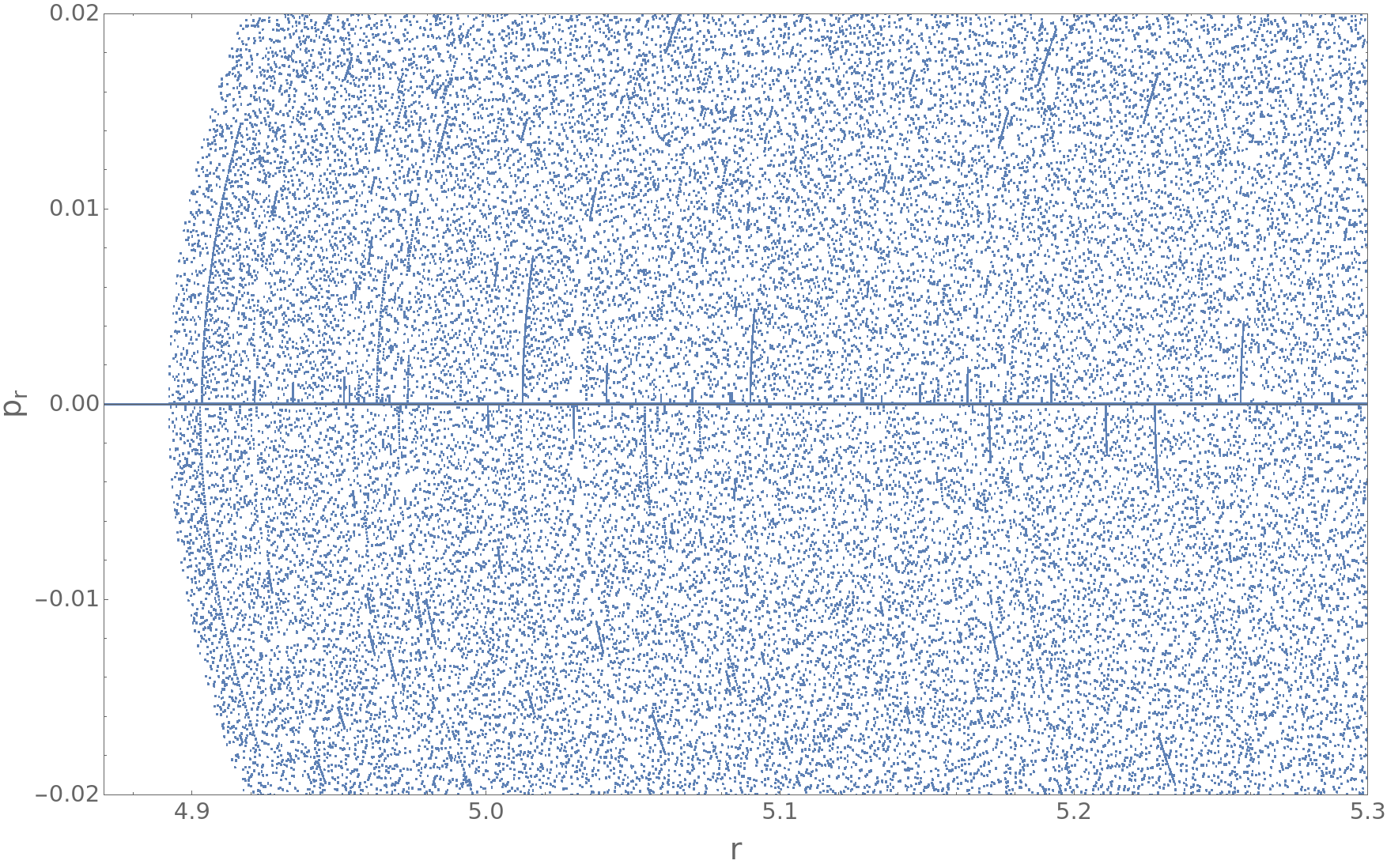}
    \caption{Poincar\'e section using HTdip metric for $\theta = \pi/2$ and $p_\theta \geq 0$ and parameters $M = 1.0$, $a = 0.1$, $E = 0.95$, $L_z = 3.0$, $\mu_{d} = 0.2$, $q_{KL} = 0.5$.}
    \label{fig:HTdipmu02q05}
\end{figure}

Figures~\ref{fig:KLdipmu02q05} and \ref{fig:HTdipmu02q05} show the combined effect of the magnetic dipole moment $\mu_d = 0.2$ and the mass quadrupole moment $q_{KL} = 0.5$ on the phase space. In both metrics, the structures besides the main island of stability are no longer visible; even chaotic regions disappear. It is not claimed that this particular combination of parameters gives rise to an integrable system, but at least the less stable geodesics are destroyed, giving the appearance of an integrable dynamical system. The main difference between both cases is the starting point of the main island of stability, which is slightly shifted in HTdip with respect to KLdip.

\begin{figure}[h]
    \centering
    \includegraphics[width=0.9\linewidth]{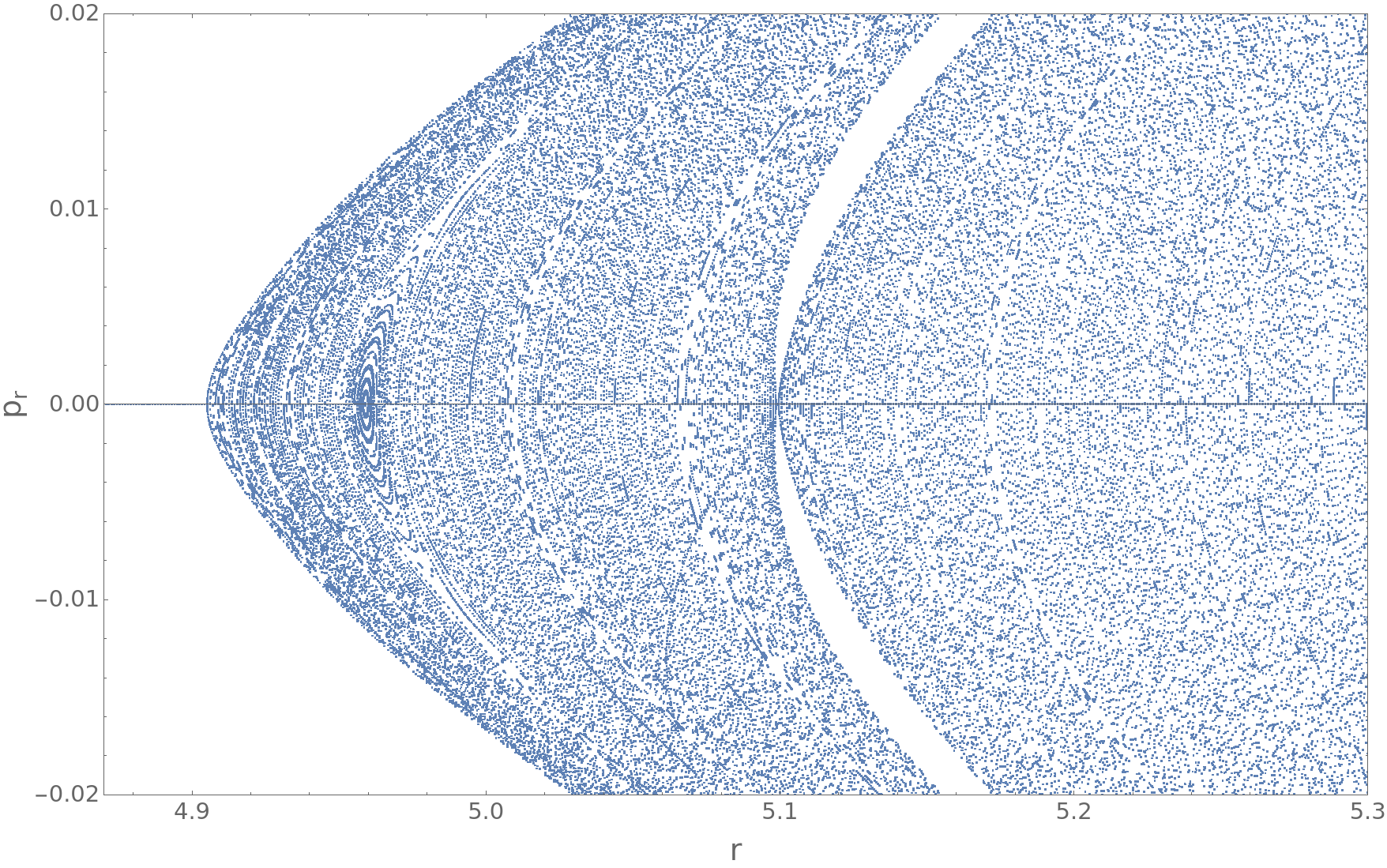}
    \caption{Poincar\'e section using KLdip metric for $\theta = \pi/2$ and $p_\theta \geq 0$ and parameters $M = 1.0$, $a = 0.1$, $E = 0.95$, $L_z = 3.0$, $\mu_{d} = 0.2$ and $q_{KL} = -0.5$.}
    \label{fig:KLdipmu02q-05}
\end{figure}

\begin{figure}[h]
    \centering
    \includegraphics[width=0.9\linewidth]{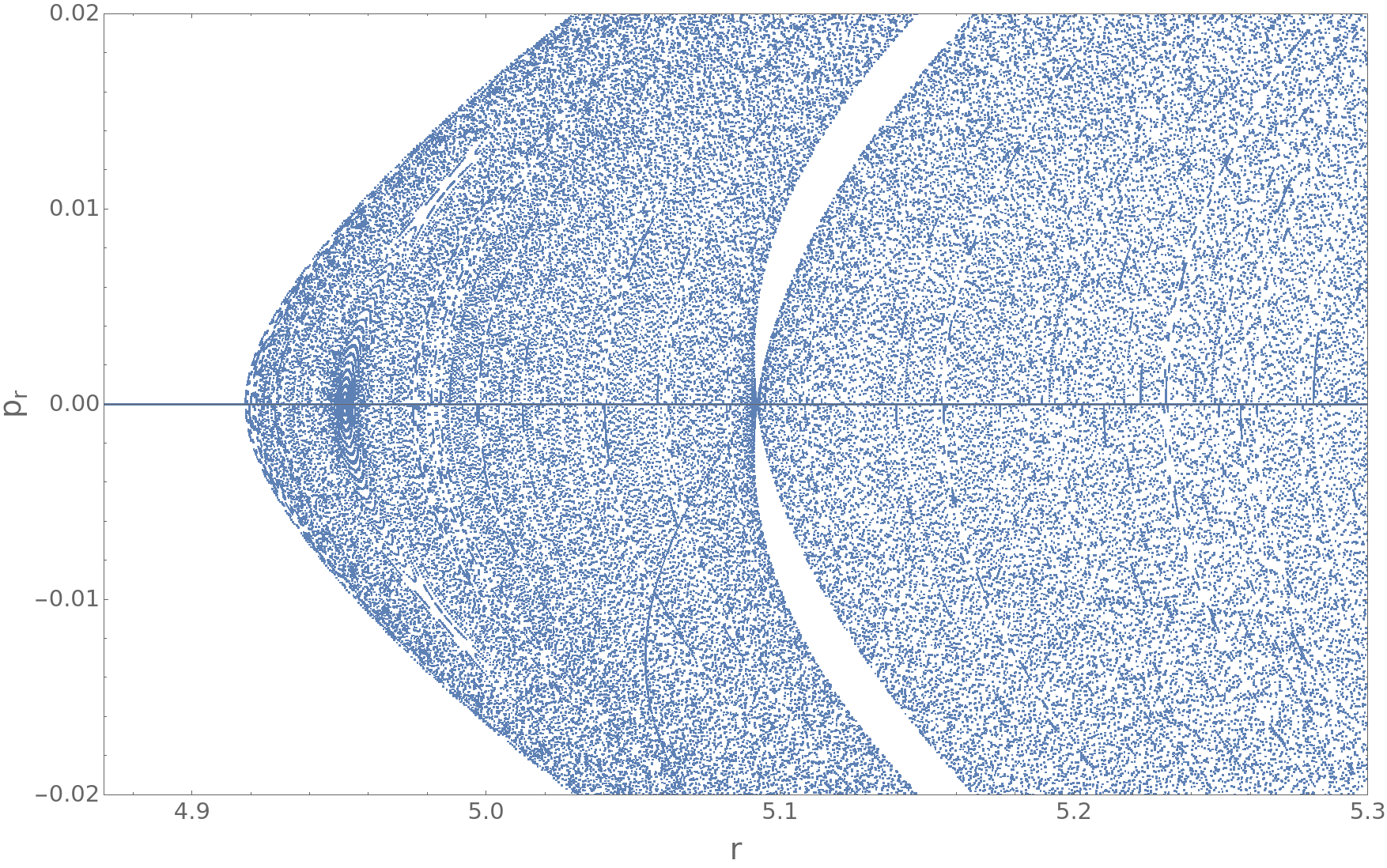}
    \caption{Poincar\'e section using HTdip metric for $\theta = \pi/2$ and $p_\theta \geq 0$ and parameters $M = 1.0$, $a = 0.1$, $E = 0.95$, $L_z = 3.0$, $\mu_{d} = 0.2$, $q_{KL} = -0.5$.}
    \label{fig:HTdipmu02q-05}
\end{figure}


Figures~\ref{fig:KLdipmu02q-05} and \ref{fig:HTdipmu02q-05} show the case of an oblate object with $\mu_d = 0.2$ and $q_{KL} = -0.5$. Again, there are no visible traces of chaos, but both sections show the same structures near $r = 4.96$ and a hyperbolic point near $r = 5.1$.

\section{Conclusions and Final Remarks}

As astronomical observations indicate, nearly every object in the universe rotates around some axis. These spinning objects deform according to their angular velocity because their equilibrium condition cannot maintain a perfect spherical shape, except the special case of black holes. For this reason, the Kerr metric is no longer valid, since it models the spacetime surrounding a perfect sphere. The dynamical system of a test particle orbiting in a Kerr spacetime is integrable; therefore, no chaos is present. However, non-Kerr systems lose their Carter constant and may show chaotic regions near their event horizon.

Both the Kerr-like metric and the Hartle--Thorne metrics represent the spacetime surrounding a spheroidal object; the deformation is given by their respective mass quadrupole moment. This parameter is used to measure the strength of the deviations from the integrable system. Mathematically, the three metrics are related to the unperturbed Kerr metric, and all three even present an \emph{event horizon}, although this is only a mathematical construct. Classically, black holes (and the hypothetical white holes) are objects described only by three parameters: mass $M$, electric charge $q_e$, and spin parameter $a$. Moreover, they have exterior spherical symmetry; hence, the possible metrics to describe them are the exact Schwarzschild ($a = q_e = 0$), Reissner--Nordstr\"om ($a = 0$), Kerr ($q_e = 0$), and the most general Kerr--Newman ($a \ne 0$, $q_e \ne 0$). However, the rest of the astrophysical objects in the universe are described by additional parameters, since they may be deformed (mass multipole moments) and may also present electromagnetic fields (electric and magnetic multipole moments). For these objects, the location of the event horizon is solely a mathematical construct and does not correspond to a physical one.

Although only the Kerr-like metric is reduced exactly to Kerr when $q_{KL} = 0$, the Hartle--Thorne metrics are approximately reduced to Kerr. This has an impact on the phase space of both HT metrics when $q_{KL} = 0$.

The behavior of the phase space between oblate and prolate objects is similar; in all three cases, chaotic regions appear at the left tip of the main island of stability. Visually, they exhibit an island surrounded by satellites, where the last stable orbit was once located.

The Geroch--Hansen multipoles are different for both metrics, especially $\mathcal{M}_2$, which involves different kinds of mass quadrupole moment; however, for the KL and HT metrics their $\mathcal{M}_2$ moments are related by equation~(\ref{E:rel_q}). The structures in the Poincar\'e sections for both prolate and oblate objects show that the KL and HTlog metrics are divergent. The difference in the behavior of the phase space is small for $q_{KL} = 0$, but it increases for higher $|q_{KL}|$. Nevertheless, the appHT and KL metrics reproduce similar results even for high $|q_{KL}|$.

The dynamical information recorded in the Poincar\'e sections contains all the information about the dynamics of the system corresponding to a test particle orbiting a compact object \cite{adrian}. As previously stated, the Hartle--Thorne metrics and the Kerr-like metric with mass quadrupole moment are related through a Taylor expansion and by comparison of their multipole moments. The previous simulations show that there is a difference between the classical Hartle--Thorne (HTlog) and the Kerr-like dynamical systems, even when both represent the same spacetime. Nevertheless, the approximate Hartle--Thorne metric lacks logarithmic terms and instead is expressed with exponential factors, which are more suitable for numerical computations and analysis. Therefore, Poincar\'e sections are powerful tools used to test the kinds of divergences that may present along different solutions of the EFE.

These systems were integrated using the Runge--Kutta--Fehlberg method with the same parameters for the source mass, test particle mass, orbital energy, orbital angular momentum, and spin parameter. The mass quadrupole moment $q_{KL}$ was used as the control and perturbation parameter. It is expected that the Kerr-like metric is more accurate than the logarithmic Hartle--Thorne metric because of the additional multipole moments and the higher-order terms in the Taylor expansions. These differences are expected due to the higher order of the Geroch--Hansen multipole moments of the KL metric. In addition, the range of validity of the KL metric with respect to the rotation of the source is broader than that of HTlog. The range of accuracy for the mass quadrupole moment is also broader in the KL and appHT metrics than in HTlog. It is concluded that the appHT and KL dynamical systems are well behaved, even though they are not isometric.

The combined effect of the mass quadrupole moment and the magnetic dipole moment gives rise to a dynamical system in which most of the less stable geodesics are destroyed. The number of structures appearing in the Poincar\'e sections is greatly reduced, to the point that there may be none besides the main island of stability. Although not shown in this paper, additional cases were simulated and exhibited the same behavior arising from the interaction between $\mu_d$ and $q_{KL}$. The implications of this result are of great importance. Spinning and slightly deformed compact objects with strong magnetic fields may deform the surrounding spacetime in such a way that satellites could exist in stable orbits. This may help explain, in part, the existence of planets orbiting pulsars. The selected parameters do not correspond to a realistic neutron star, since large values of $|q_{KL}|$ represent a highly flattened or elongated object. These parameters, especially $q_{KL}$ and $\mu_d$, were chosen to reveal as many phase-space structures as possible and to test the applicability of the metrics.

Finally, the higher validity of the multipole moments of the KL metric compared to its HT counterparts, and the same occurring for KLdip with respect to HTdip, makes the KL and KLdip metrics better suited for numerical simulations and for testing general relativity in applications to real observations.


\end{document}